\documentclass{aa}  

\usepackage{xcolor}
\usepackage{longtable}
\usepackage{amsmath,amssymb}
\usepackage{graphicx}
\usepackage{url}
\usepackage{color}

\newcommand{\Rmnum}[1]{\uppercase\expandafter{\romannumeral #1}}
\usepackage{txfonts}
\usepackage[colorlinks=true,linkcolor=blue,citecolor=blue,urlcolor=blue]{hyperref}
\usepackage{natbib,twoopt}
\begin{document}

   \title{Likely detection of magnetic field related LFQPO in the soft X-ray re-brightening of GRS~1915+105}


   \author{Ling-Da Kong\inst{1}
          \and
          Long Ji\inst{2}
          \and
          Andrea Santangelo\inst{1}
          \and 
          Meng-Lei Zhou\inst{1}
          \and 
          Qing-Cang Shui\inst{3,4}
          \and 
          Shu Zhang\inst{3}
          }

   \institute{Institut f{\"u}r Astronomie und Astrophysik, Kepler Center for Astro and Particle Physics, Eberhard Karls, Universit{\"a}t, Sand 1, D-72076 T{\"u}bingen, Germany\\
              \email{lingda.kong@mnf.uni-tuebingen.de}
         \and
             School of Physics and Astronomy, Sun Yat-Sen University, Zhuhai, 519082, China\\
            \email{jilong@mail.sysu.edu.cn}
        \and
            Key Laboratory for Particle Astrophysics, Institute of High Energy Physics, Chinese Academy of Sciences, 19B Yuquan Road, Beijing 100049, China
        \and
            University of Chinese Academy of Sciences, Chinese Academy of Sciences, Beijing 100049, China
             }


 
\abstract
    {Utilizing NICER observations, we present an analysis of the soft X-ray re-brightening event of GRS 1915+105 observed in 2021. During this event, we observed the emergence of a stable, long-lasting low-frequency quasi-periodic oscillation (LFQPO) with frequencies ranging from 0.17 to 0.21 Hz. Through a careful spectral analysis, we demonstrate that a low-temperature Compton-thick gas model well characterizes the emitted radiation. By examining the spectrum and identifying numerous absorption lines, we discerned a transition in the wind properties. This transition was marked by a shift from a state characterized by low speed, high column density, and high ionization degree to one featuring still low speed but low column density and ionization degree. Intriguingly, the presence or absence of the QPO signal is perfectly correlated with these distinct wind characteristics. The low-speed wind observed could be indicative of a 'failed wind', while the observed shift implies a transition from a magnetically to a thermally driven wind. Notably, this QPO signal exclusively manifested itself during the magnetically driven phase, suggesting the possibility of a novel perturbation associated with magnetic effects.}

\keywords{accretion, accretion disks -- stars -- black hole -- stars -- magnetic field -- X-rays: binaries}
\maketitle
%

\section{Introduction}

GRS~1915+105, a microquasar discovered in 1992 by the GRANAT/WATCH all-sky monitor \citep{1992IAUC.5590....2C}, is well known for its remarkable variability.
It consists of a 12 $M_{\odot}$, high spin BH and a K-M \Rmnum{3} companion star (\citealp{2001A&A...373L..37G, Reid2014ApJ...796....2R}).  
Since its discovery, GRS~1915+105 has been active for 26 years and has a long history of being monitored by all major all-sky X-ray telescopes.
During this long-term period, its variability has been classified into at least 15 different states with unique timing-spectral properties (\citealp{2000A&A...355..271B, 2002MNRAS.331..745K, Hannikainen2005A&A...435..995H, Athulya2022MNRAS.510.3019A}).
In 2018, experiencing an exponential decay during MJD 58230-58330 and a linear decay during MJD 58330-58500, the source was in a low flux level (see \citealp{Koljonen2020, Koljonen2021A&A...647A.173K} for details).
Still in the radio, IR, and X-ray bands, significant variability was observed (\citealp{Murata2019, Trushkin2019, Koljonen2019, Trushkin2019b, Trushkin2020, Vishal2019, Murata2019, Motta2021MNRAS.503..152M, Egron2023ATel16008....1E, Sanchez-Sierras2023ATel16039....1S}).   
GRS~1915+105 seems to have transitioned from a canonical hard state to an unusual accretion phase characterized by heavy X-ray absorption/obscuration. The presence of a Compton-thick 'failed wind'  (\citealp{Miller2020ApJ...904...30M, Balakrishnan2021ApJ...909...41B}), which hides from the observer the accretion processes feeding the variable jet responsible for the radio flaring has been discussed (\citealp{Motta2021MNRAS.503..152M}). 
This kind of obstruction may be unstable, allowing the internal accretion processes and jets to be briefly observed.
In addition, the long-term low X-ray level is always interrupted by short X-ray flares with synchronized radio flares within a few hundred seconds (\citealp{Trushkin2019, Homan2019ATel13308....1H, Neilsen2020ApJ...902..152N, Kong2021ApJ...906L...2K, Trushkin2023ATel15964....1T, Egron2023ATel16008....1E}).

In the case of GRS~1915+105, it has been observed that massive winds are predominantly found in softer states, although not exclusively (\citealp{Miller2020ApJ...904...30M, Ratheesh2021, Neilsen2020ApJ...902..152N, Kong2021ApJ...906L...2K}).
During these softer states, strong jet emissions are generally absent or relatively weak (\citealp{Ponti2012, Homan2016}).
\cite{Neilsen2009} have proposed that these massive winds significantly influence the disk accretion flow. 
They could suppress the formation of jets or even completely quench the jet activity in the system.
However, in some exceptional cases, a rapid transition between the jet and the wind can occur in the heartbeat state ($\rho$ state).
As discussed by \cite{2011ApJ...737...69N}, the jet on smaller scales during a short period with a 10\% cycle near the minimum luminosity of the pulse of the heartbeat, and the disk wind at larger scales may lead the fast spectral transitions in different phases.
\cite{Zoghbi2016} also found the presence of two wind components with velocities between 500 and 5000 km/s and possibly two more with velocities reaching 20,000 km/s ($\sim$0.06 c).
The alterations in wind speed and absorbed flux necessitate rapid changes in wind geometry and location within the heartbeat cycle.

Such an evolution of the wind features may be associated with a bulge, born in the inner disk and moving outward as the instability progresses (\citealp{Zoghbi2016}).
Through a broad-band spectral analysis of the X-ray/radio flare, \cite{Kong2021ApJ...906L...2K} found a fast transition between jet and magnetically driven wind.
All these results imply that GRS~1915+105 can be a laboratory to study the relations among jet, wind, and magnetic fields on the accretion disk.

In July 2021, a soft X-ray re-brightening was monitored by MAXI/GSC, NICER, and AstroSat(\citealp{Neilsen2021ATel14792....1N, Ravishankar2021ATel14811....1R}), which is the second re-brightening in the low-flux state after 2018.
During this re-brightening, \cite{Neilsen2021ATel14792....1N} found that the spectrum contains over a dozen strong absorption lines from Si, S, Ar, Ca, Fe, and Ni, as well as an emission line consistent with Fe K-alpha. This suggests the presence of an ionized absorber with $N_{\rm h}>5\times10^{23}$ cm$^{-2}$ without evidence of cold Compton-thick obscuration. 
Meanwhile, as the intensity increased, they found a low-frequency QPO (LFQPO) at 0.17 Hz.
In black hole binary systems (BHXRBs), low-frequency quasi-periodic oscillations (LFQPOs) typically occur in the Low Hard State (LHS) and High Intermediate State (HIMS), while the disk wind is commonly observed in the high soft State (HSS). 
In this soft X-ray re-brightening, the simultaneous presence of both multiple absorption lines (wind feature) and LFQPOs is a truly rare and intriguing event. Through such observations, we can further investigate the nature of disk winds and the origin of LFQPOs.

In this article, we focus on an analysis of the evolution of spectral and timing variations through the long-lasting soft X-ray re-brightening of 2021. We also focus on the physical properties and origin of the absorber and LFQPOs. In this letter, we introduce the data reduction and analysis in Section 2 and present the results in Section 3. Finally, we discuss this in Section 4, and a conclusion is given in Section 5.
\section{Observations and Data reduction}

The NICER X-ray Timing Instrument (XTI) comprises an array of 56 co-aligned concentrator X-ray optics, which are paired with a single-pixel silicon drift detector working in the 0.2–12 keV band (\citealp{Gendreau2016SPIE.9905E..1HG}). NICER's energy and time resolutions are 85 eV at 1 keV and 40 ns, respectively.

In this work, we use NICER observations between June 17, 2021 (MJD 59382) and October 3, 2021 (MJD 59490), performed during the soft X-ray re-brightening of the source. Due to the occasional increase of the electronic noise, detectors No. 14 and 34 are excluded from the analysis.

NICER is an excellent telescope for observing soft X-rays due to its effective area of $\sim 1800$ cm$^2$ at 1.5 keV. We have used the \texttt{nicerl2} pipeline task to process each observation by applying the default calibration process and screening. The spectra and light curves have been extracted by the \texttt{nicerl3} pipeline task. The background is calculated with the ``nibackgen3C50'' tool provided by the NICER team. The data analysis is under the \texttt{Heasoft\ v6.31.1} environment, and the NICER's calibration is updated to October 1, 2022.

In this work, we also use the NuSTAR ToO observation started on July 14, 2021, at 06:36:09 (UTC), with a total exposure of 20.44 ks (ObsID: 90701323002).
The NuSTAR Data Analysis Software (\texttt{NuSTARDAS}) including \texttt{nupipeline v0.4.9} and \texttt{nuproducts} with the calibration database version v20230420 is used in this work. We extract the NuSTAR lightcurves of the source and the background with two circular extraction regions of $100''$ and $180''$, one centered at the point-like source and another centered away from the source.

All spectra are grouped by \texttt{ftgrouppha} task using the \cite{Kaastra2016A&A...587A.151K} optimal binning scheme with a minimum of 25 counts per bin.
Our timing analysis is based on the Python package \texttt{Stingray} (\citealp{Huppenkothen2019JOSS....4.1393H,Huppenkothen2019ApJ...881...39H}) and on Xronos 6.0 task \texttt{powspec} (\citealp{Stella1992ASPC...25..103S}).
We use Xspec v12.13.0c (\citealp{1996ASPC..101...17A}) to study the spectrum in the frequency and energy domain.
We choose the energy range $1-10$ keV of NICER and $4-25$ keV of NuSTAR for spectral analysis and take the frequency range 1/128 to 10 Hz for timing analysis.
The uncertainties of the spectral parameters are computed using the Markov Chain Monte Carlo (MCMC) method with a length of 10,000 and are reported at a 90\% confidence level.

\section{Data analysis and observational results}
\begin{figure}
  \resizebox{\hsize}{!}{\includegraphics{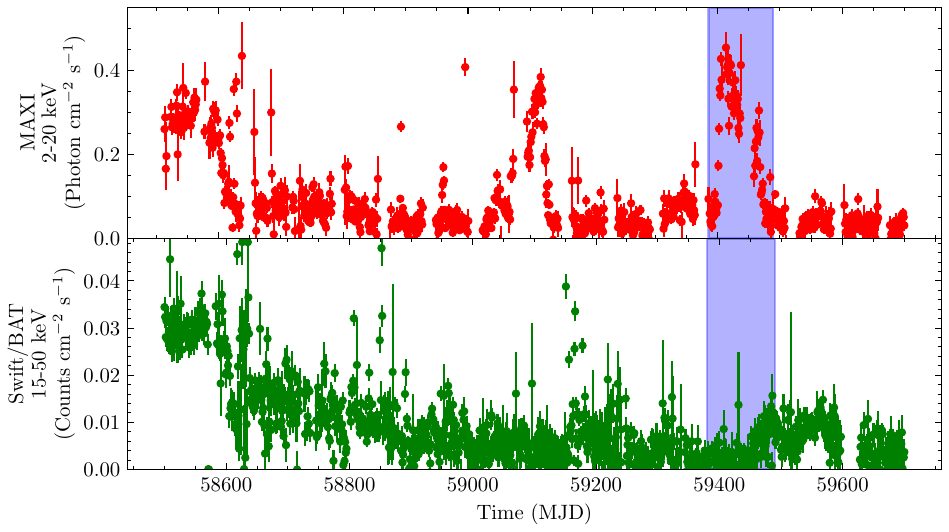}}
  \caption{The figure shows the light curves derived from the MAXI all-sky monitor in the $2-20$ keV range and Swift/BAT in the $15-50$ keV range. The source transition into an extended low flux state is clearly observed. The noteworthy resurgence of brightness in 2021, which is the primary focus of our study, is indicated in blue.}
  \label{maxi-bat}
\end{figure}

\begin{figure}
    \resizebox{\hsize}{!}{\includegraphics{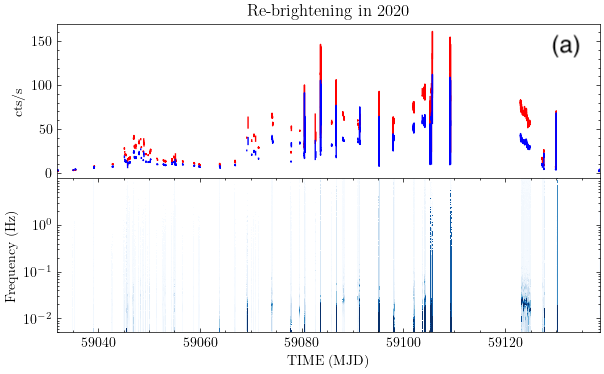}}
    \resizebox{\hsize}{!}{\includegraphics{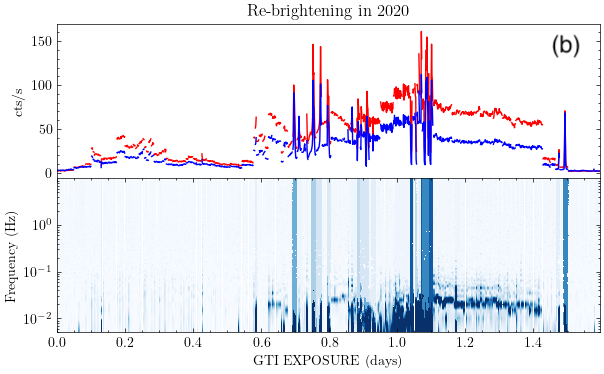}}
    \caption{(a): The top panel displays the re-brightening in 2020 in the $1-4$ keV and $4-10$ keV, represented by red and blue curves, respectively. In the bottom panel, we present the dynamic power density spectrum (PDS) spanning the $1-10$ keV range. (b): We have eliminated the gaps between effective observations due to the extended time scale of the data to enhance the clarity of the temporal evolution. Notably, the figure clearly reveals the presence of mHz QPOs $\sim20$ mHz, a pattern that aligns with the heartbeat state.}
    \label{DPS_2020}
\end{figure}

\begin{figure}
    \resizebox{\hsize}{!}{\includegraphics{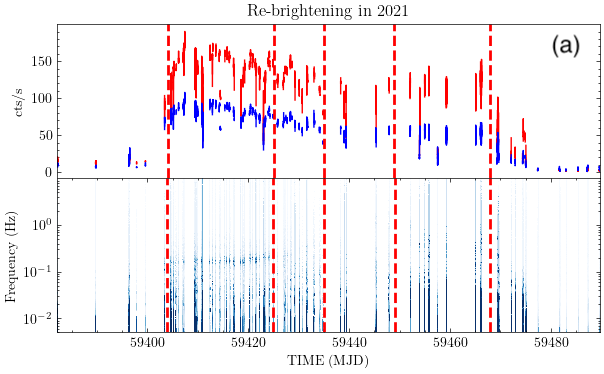}}
    \resizebox{\hsize}{!}{\includegraphics{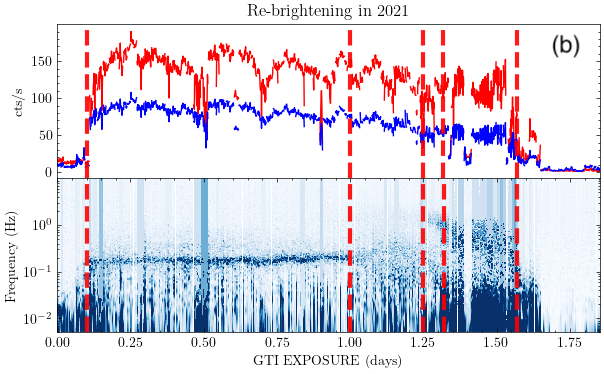}}
    \caption{(a): The top panel displays the re-brightening in 2021 in the $1-4$ keV and $4-10$ keV, represented by red and blue curves, respectively. In the bottom panel, we present the dynamic power density spectrum spanning the $1-10$ keV range. The five red dashed lines serve to delineate four distinct epochs characterized by different timing properties. (b): We have eliminated the gaps between effective observations due to the extended time scale of the data to enhance the clarity of the temporal evolution. Notably, the figure clearly reveals the presence of LFQPOs $\sim0.2$ Hz and high-frequency brake noise. The five dashed lines in (a) and (b) represent the same time intervals. With the gaps removed, they more clearly distinguish four different regions in (b).}
    \label{DPS_2021}
\end{figure}

In Figure~\ref{maxi-bat}, we present the light curves from the all-sky monitor MAXI in the $2-20$ keV range and from Swift/BAT in the $15-50$ keV range after the source enters the low flux state.  
In 2020 and 2021, the source experienced two episodes of soft re-brightening, both primarily characterized by an abundance of soft photons.
In the top panel of Figure~\ref{DPS_2020} and Figure~\ref{DPS_2021}, we present NICER light curves in the $1-4$ keV (red points) and $4-10$ keV (blue points). In the bottom panel, we display the dynamic power spectrum (DPS) in the $1-10$ keV band, which reveals significant variability. 
Each power density spectrum is computed using Leahy normalisation (\citealp{Leahy1983ApJ}), then we subtracted the Poisson noise that in the case of this normalisation levels at 2, to highlight the physical signals (see Figure~\ref{DPS_2020}(a) and Figure~\ref{DPS_2021}(a)).
Due to the long time span of the observation, we have removed the gaps between effective observational windows to show more clearly the evolution (see Figure~\ref{DPS_2020}(b) and Figure~\ref{DPS_2021}(b)).
The GTI EXPOSURE (days) means each observation's net effective exposure time. These time slices are sorted in chronological order of observations. The X-axis in Figure~\ref{DPS_2020}(b) and Figure~\ref{DPS_2021}(b) shows the total GTI accumulation as time evolves.
For the first re-brightening, there is no finding of any LFQPO in $0.1-10$ Hz, but there are mHz QPOs with $\sim20$ mHz whose light curve can be classified into the heartbeat state.
The primary emphasis of this paper is directed towards the second re-brightening since it shows a more stable LFQPO (at $\sim0.2$ Hz) when compared to the first re-brightening.
The second re-brightening exhibits a sharp flux increase and a slowly decreasing trend with a few sudden dips.
In the bottom of Figure~\ref{DPS_2021}, we can discern four distinct epochs that characterize the re-brightening: 1) QPO dominated at $\sim 0.17$ Hz (Epoch 1, GTI EXPOSURE $0.1-1.0$ days); 2) low-frequency noise dominated at $\sim$ 1 Hz (Epoch 2, $1.0-1.25$ days); 3) peaked noise dominated at $\sim 1.5$ Hz (Epoch 3, $1.25-1.32$ days); and 4) high-frequency noise dominated at$\sim 1$ Hz (Epoch 4, $1.32-1.57$ days). 
It's important to note that the power spectral density during the dip periods is predominantly influenced by Poisson noise. 
For our spectral and timing analysis, we have selected 41 observations based on two criteria: each observation has to have an exposure time of at least 500 seconds and a count rate exceeding 100 cts/s to ensure statistical significance.

\subsection{Timing variability}

\subsubsection{Power density spectrum analysis}
\begin{figure*}
    \centering\includegraphics[width=0.49\textwidth]{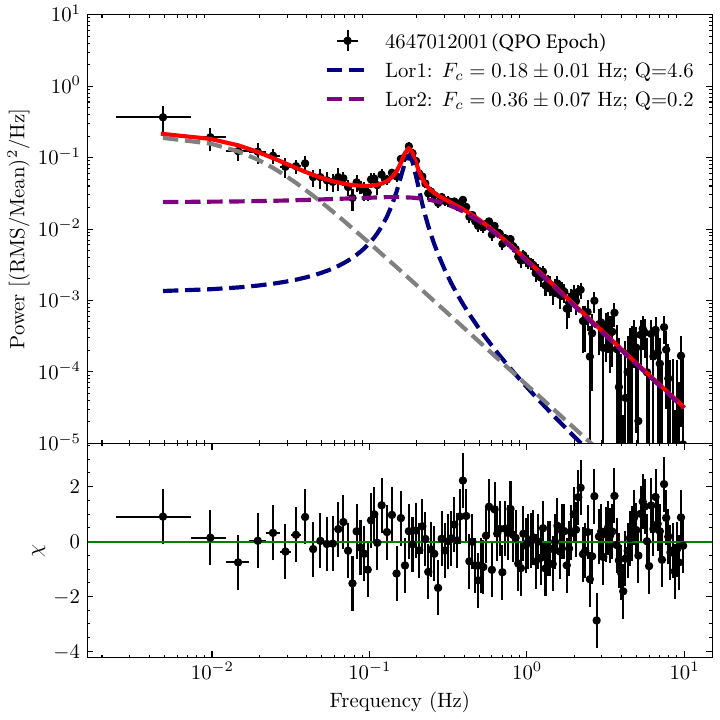}
    \centering\includegraphics[width=0.49\textwidth]{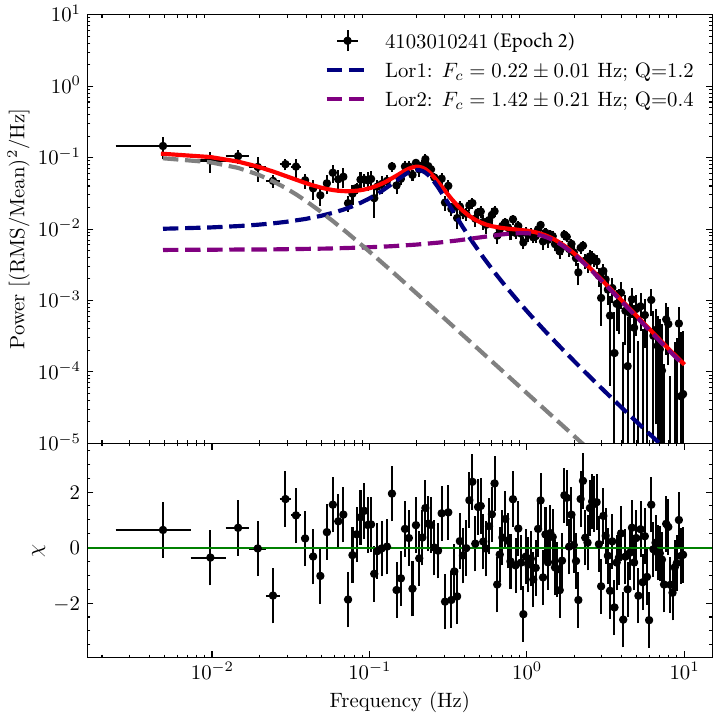}
    \centering\includegraphics[width=0.49\textwidth]{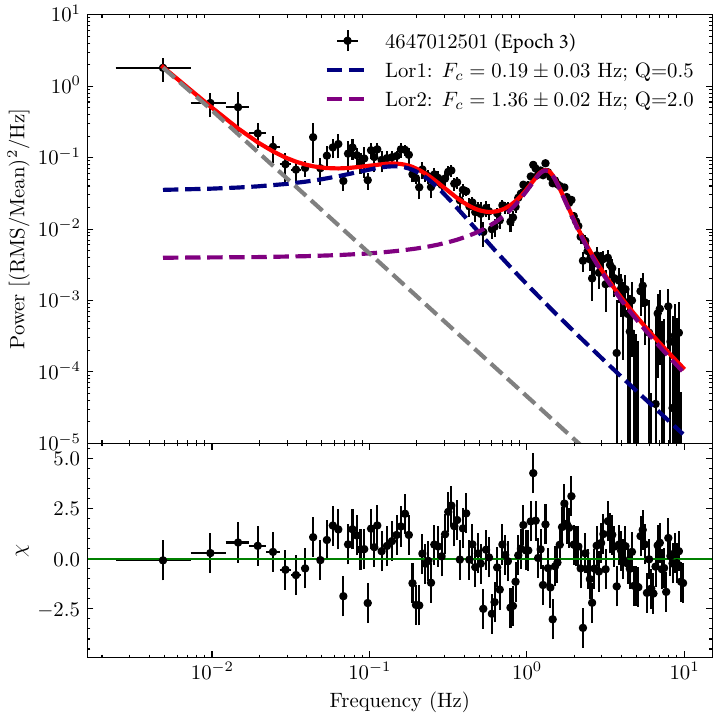}
    \centering\includegraphics[width=0.49\textwidth]{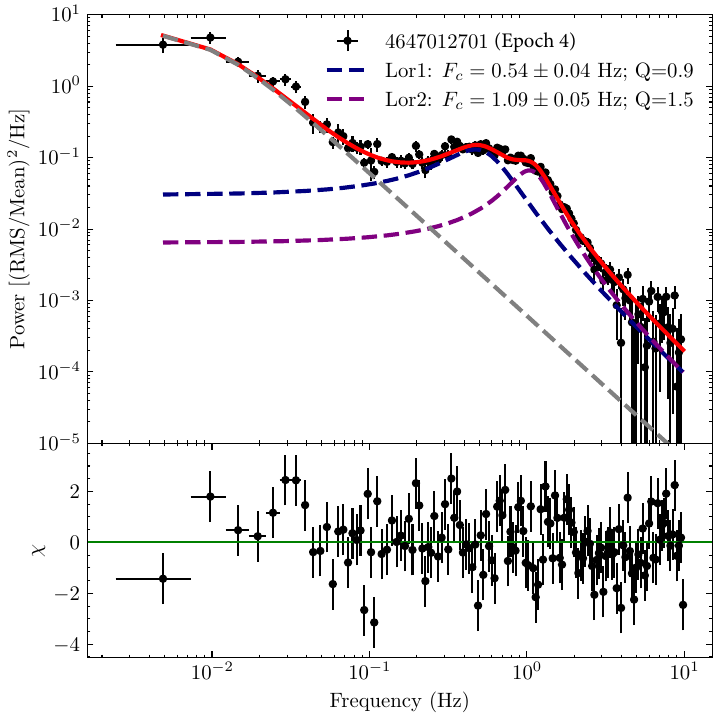}
    \caption{The PDSs of ObsID 4647012001 (QPO Epoch), 4103010241 (Epoch 2), 4647012501 (Epoch 3), and 4647012701 (Epoch 4) are fitted with three Lorentzian components shown in blue (Lor1), purple (Lor2), and gray dashed lines. The red line indicates the sum of all components. The bottom part of each panel shows the residual of the fitting. The characteristic frequency $F_{\rm c}$ and the Q factor of the Lor1 and Lor2 components are shown in each panel.}
    \label{PDS_fitting}
\end{figure*}

An in-depth timing analysis of the 41 observations has been performed. For each of these observations, we employ the \texttt{powspec} tool to generate a power density spectrum (PDS) using the screened event file within the frequency range of $1/128-10$ Hz and in the energy band of $1-10$ keV.
The PDS is converted to a readable file for Xspec by the \texttt{flx2rsp} task, and we use multi-Lorentz components to fit the broadband, the peaked noises, and QPOs.
For each of the four epochs mentioned earlier, we identify one observation to represent its timing properties.
To show the details of their PDS fit (see Figure~\ref{PDS_fitting}) we choose ObsID 4647012001 for Epoch 1, ObsID 4103010241 for Epoch 2, ObsID 4647012501 for Epoch 3, and ObsID 4647012701 for Epoch 4.
Each observation can be accurately fitted with three Lorentzian components, which we have depicted using dashed lines in blue (Lor1), purple (Lor2), and gray (Lor3).
Regarding the Lor3 component, we have fixed the central frequency ($F_0$) at 0, as it consistently represents band-limited noise that prevails below 0.1 Hz throughout the entire re-brightening period.
However, it's worth noting that the other two components exhibit distinct properties across different epochs. For these two components, we have maintained the flexibility to adjust all parameters during the fitting process.
We use the characteristic frequency $F_{\rm c}\equiv F_0\sqrt{1+1/(4Q^2)}$, at which it attains the maximum power in $\nu P_{\nu}$ and Q factor $Q\equiv F_0/\sigma$ to measure the narrowness of the component. The $\sigma$ is the full width at half maximum (FWHM) of the Lorentz component.

In ObsID 4647012001 during Epoch1, the Lor1 appears as a narrow QPO component with $F_{\rm c}=0.18\pm0.01$ Hz and $Q=4.6$, which is larger than 2. But it becomes a wide noise component with $Q<2$ in ObsID 4103010241 during Epoch2, ObsID 4647012501 during Epoch3, and ObsID 4647012701 during Epoch4. 
The Lor2 component evolves from a band-limited noise to a peaked noise with $F_{\rm c}=1.36\pm0.02$ Hz and $Q=2.0$ in ObsID 4647012501 during Epoch3 and with $F_{\rm c}=1.09\pm0.05$ Hz and $Q=1.5$ in ObsID 4647012701 during Epoch4.

\subsubsection{Timing properties as a function of time}

\begin{figure*}
    \centering\includegraphics[width=0.49\textwidth]{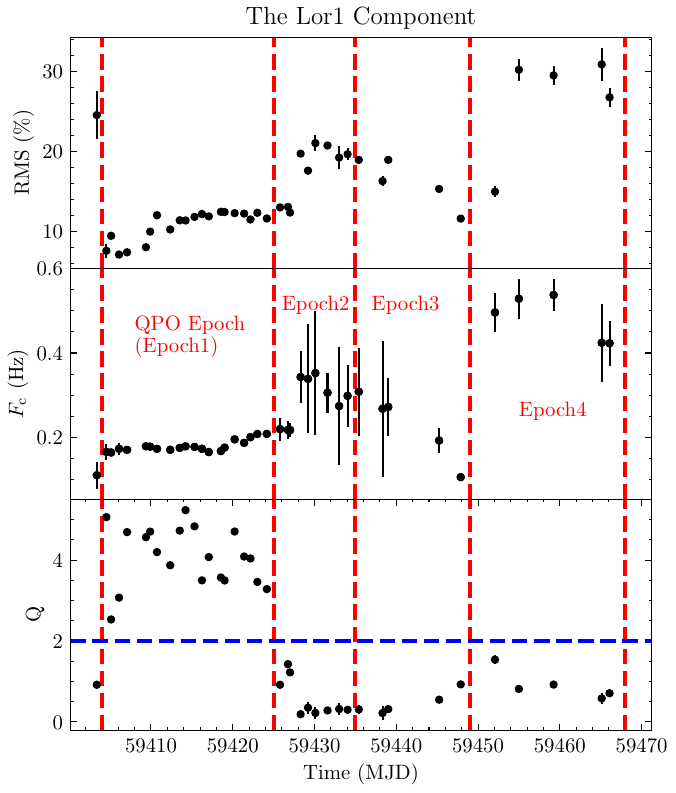}
    \centering\includegraphics[width=0.49\textwidth]{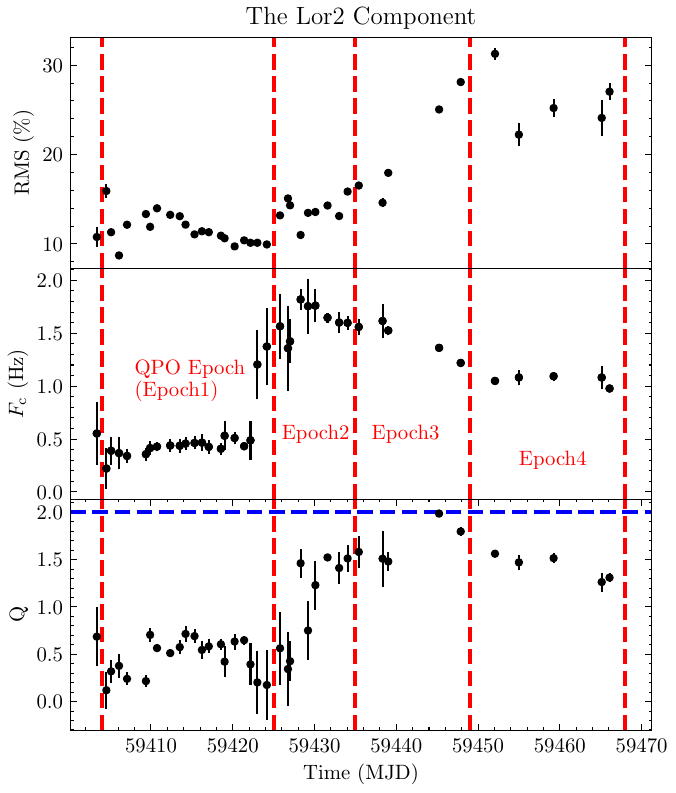}
    \caption{The RMS (\%), $F_{\rm c}$ (Hz), and Q factor of the Lor1, Lor2 are shown in the left and right panels. The five red dashed lines serve to delineate four distinct epochs characterized by different timing properties.}
    \label{Lor_evo}
\end{figure*}

In Figure~\ref{Lor_evo}, we present the temporal evolution of the Lor1 and Lor2 components in the left and right panels, respectively. In the left panel, the Lor1 component is a low-frequency QPO (LFQPO) with a central frequency ($F_{\rm c}$) in the range of approximately $0.17-0.21$ Hz. The quality factor is $Q>2$ during the period between MJD 59404 and 59425 in Epoch 1. However, after MJD 59425, it makes the transition into a band-limited noise state with $Q\ll2$.
In the right panel, during the QPO epoch, the Lor2 appears as a band-limited noise with $F_{\rm c}\sim0.4$ Hz and $Q\sim 0.5$. However, at the end of the QPO epoch, the $F_{\rm c}$ increases suddenly from $\sim0.4$ Hz to $\sim 1.5$ Hz, and the Q factor increases from $\sim 0.5$ to $\sim 1.5$. This makes the Lor2 a peaked noise or even a broad QPO with $Q\sim$ in ObsID 4647012501.

\begin{figure}
    \resizebox{\hsize}{!}{\includegraphics{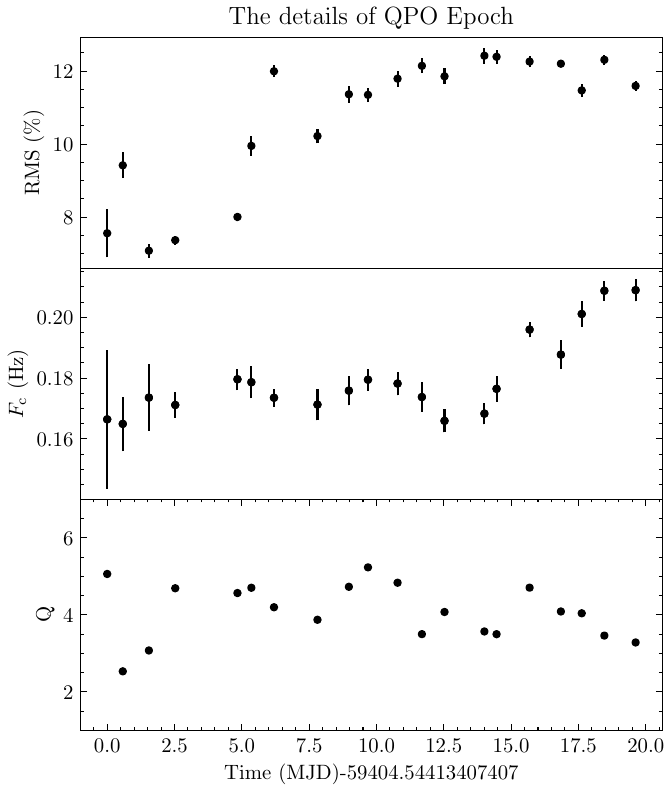}}
    \caption{The RMS (\%), $F_{\rm c}$ (Hz), and Q factor of the LFQPOs.}
    \label{qpo_evo}
\end{figure}

In Figure~\ref{qpo_evo}, we offer a more detailed view of the QPO epoch, showing its evolution over time.
We observe a gradual increase in the QPO frequency from 0.17 Hz to 0.21 Hz, with two exceptions during the dips around MJD 59412 and 59418. Additionally, the QPO's root mean square (RMS) value shows a progressive rise over time, growing from 7\% to 12\%.

\subsubsection{The RMS spectrum}

\begin{figure}
    \resizebox{\hsize}{!}{\includegraphics[width=0.49\textwidth]{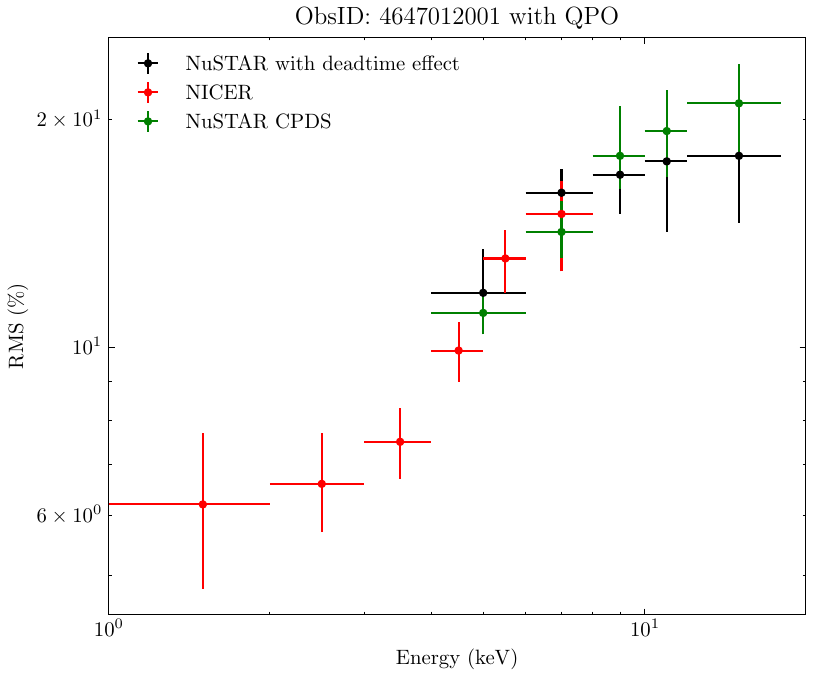}}
    {\includegraphics[width=0.49\textwidth]{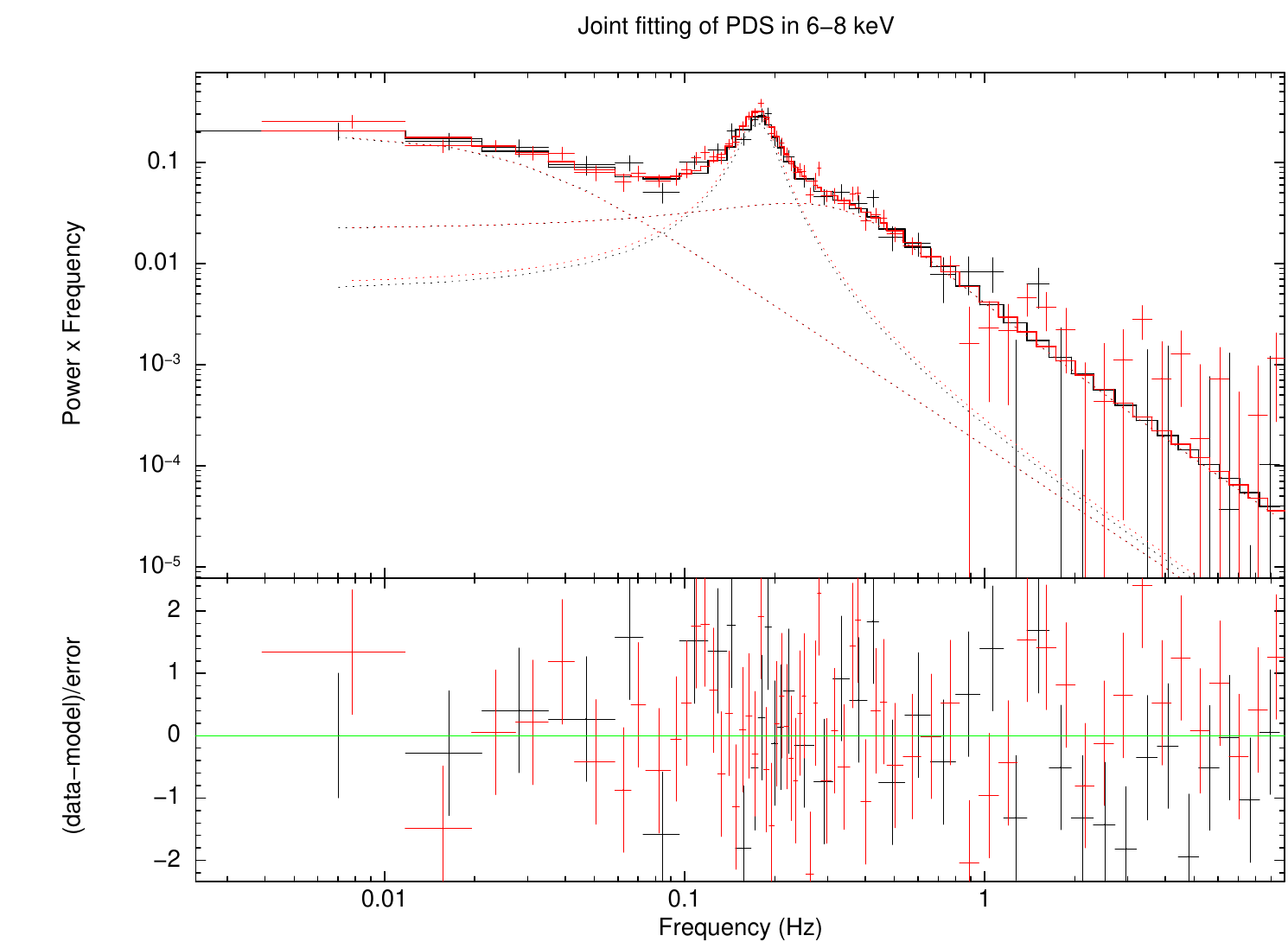}}
    \caption{Upper panel: The evolution of the RMS with energy is shown. The data points in red and black correspond to simultaneous observations conducted by NICER and NuSTAR on July 14, 2021 (NICER: 4647012001; NuSTAR: 90701323002). The green points show the RMS spectrum of NuSTAR from cross-power density spectra (CPDS). Bottom panel: The joint fitting of the PDS in $6-8$ keV. The black and red points show the PDS of NICER and CPDS of NuSTAR, separately.
    }
    \label{qpo_rms_E}
\end{figure}

To study the evolution of the QPO fractional rms with energy, we selected a joint observation of NICER and NuSTAR on July 14, 2021 (NICER: 4647012001; NuSTAR: 90701323002) within Epoch 1.
Due to the effective areas of both NICER and NuSTAR, we can obtain a significant broadband RMS spectrum. 
To account for the deadtime effect, we generated cross-power density spectra (CPDS) for NuSTAR data, using the tool \texttt{MaLTPyNT} (\citealp{Bachetti2015ApJ...800..109B}). 
This timing analysis package takes care of the NuSTAR dead time effect (\citealp{Harrison2013ApJ...770..103H, Bachetti2015ApJ...800..109B}). In order to check if this method removes the deadtime effect, we compared the DPS of NICER and CPDS of NuSTAR in $6-8$ keV in the bottom panel of Figure~\ref{qpo_rms_E}. And the RMS spectrum of NuSTAR is shown with green points in the upper panel of Figure~\ref{qpo_rms_E}.

Regarding the white noise, instead of directly subtracting it from the PDS when generating the PDS using the \texttt{powspec} tool, we opt to fit it using a \texttt{powerlaw} function with a fixed index of 0 in order to ensure that deadtime effects do not affect the calculation of QPO rms.
For the remaining components, our fitting methodology follows the same approach outlined in Section 3.1.1.

For the NICER's observation, light curves in the energy bands $1-2$ keV, $2-3$ keV, $3-4$ keV, $4-5$ keV, $5-6$ keV, and $6-8$ keV are extracted, whereas for the NuSTAR's observation, we chose $4-6$ keV, $6-8$ keV, $8-10$ keV, $10-12$ keV, and $12-18$ keV. Results are shown in the upper panel of Figure~\ref{qpo_rms_E}. From $1$ to $18$\,keV, the QPO's RMS overall increases from $6\%$ to $18\%$. More specifically, only in the band $3-8$\,keV the RMS clearly increases, while it stays stable at 6\% in the $1-3$ keV range, and around 18\% above 8\,keV.

NICER does not contain a significant QPO component above 8 keV. The PDS at $8-10$\, keV is dominated by white noise due to the low effective area at these energies and therefore limited statistics.
At higher energies ($>8$ keV), NuSTAR has a larger effective area, allowing us to constrain the QPO signal. For energies $>18$\,keV, the NuSTAR PDS is also dominated by white noise. This is not surprising since the count rate above 20 keV is very low for this outburst, and it is essentially dominated by the background.

\subsection{The spectral analysis}

\subsubsection{Joint NICER and NuSTAR spectral analysis}

The energy range of NICER does not allow us to obtain tight constraints on parameters of spectral components typically constrained by the hard energies.
To this aim, we can use the simultaneous observations by NuSTAR (ObsID: 90701323002) and NICER (ObsID: 4647012001) on July 14, 2021, in the QPO epoch. This provides us with an opportunity to validate our model through the joint fitting.
For NuSTAR's observation, limited by the background, we chose the energy range from 4 keV to 25 keV. 
During the fitting process, we included a 1.5\% systematic error for NICER. And we found wee need to add additional 1\% systematic error between $6-10$ keV for NuSTAR due to the limited spectral resolution for the absorption lines.

\begin{figure*}
    \centering\includegraphics[width=0.49\textwidth]{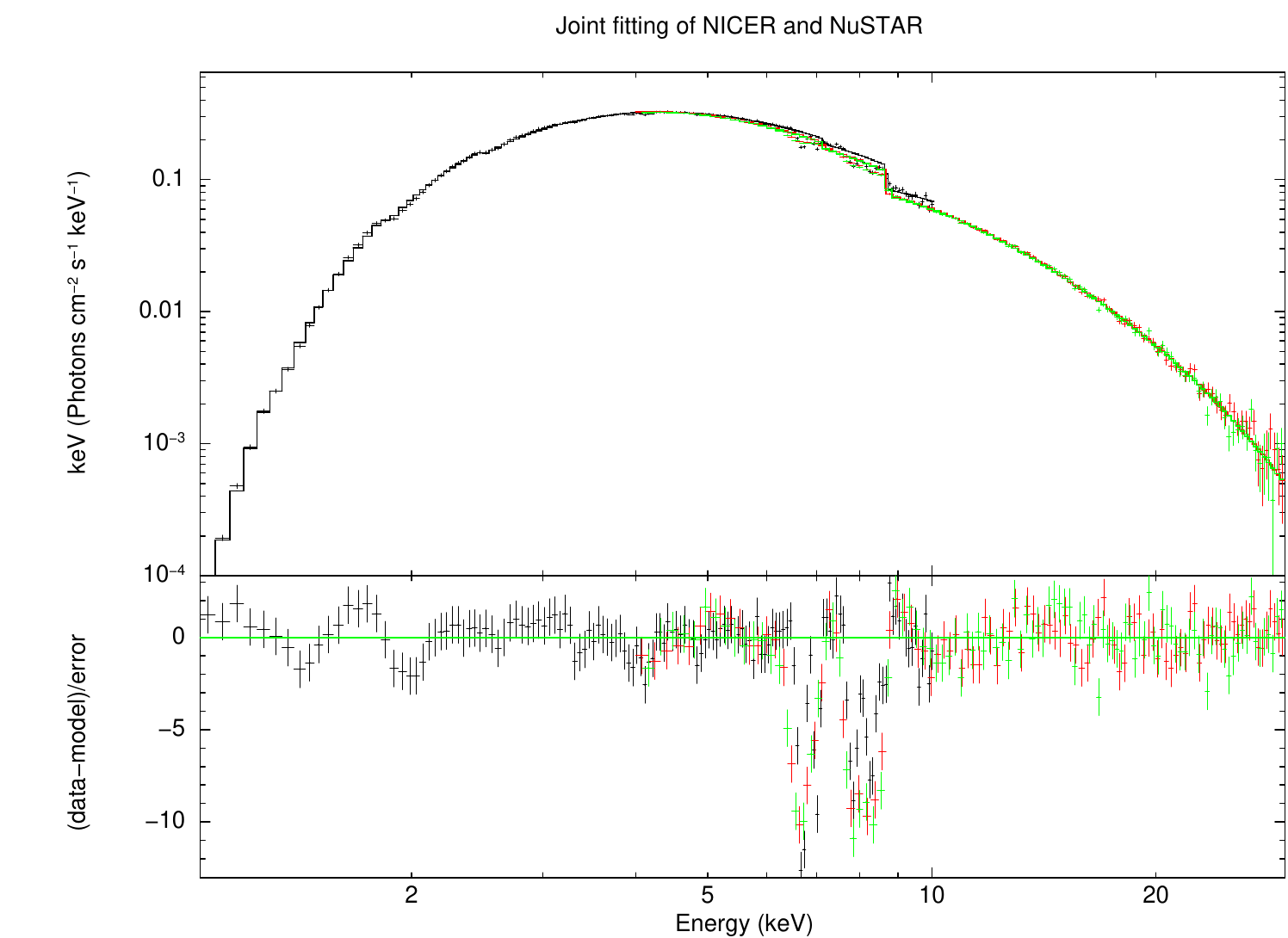}
    \centering\includegraphics[width=0.49\textwidth]{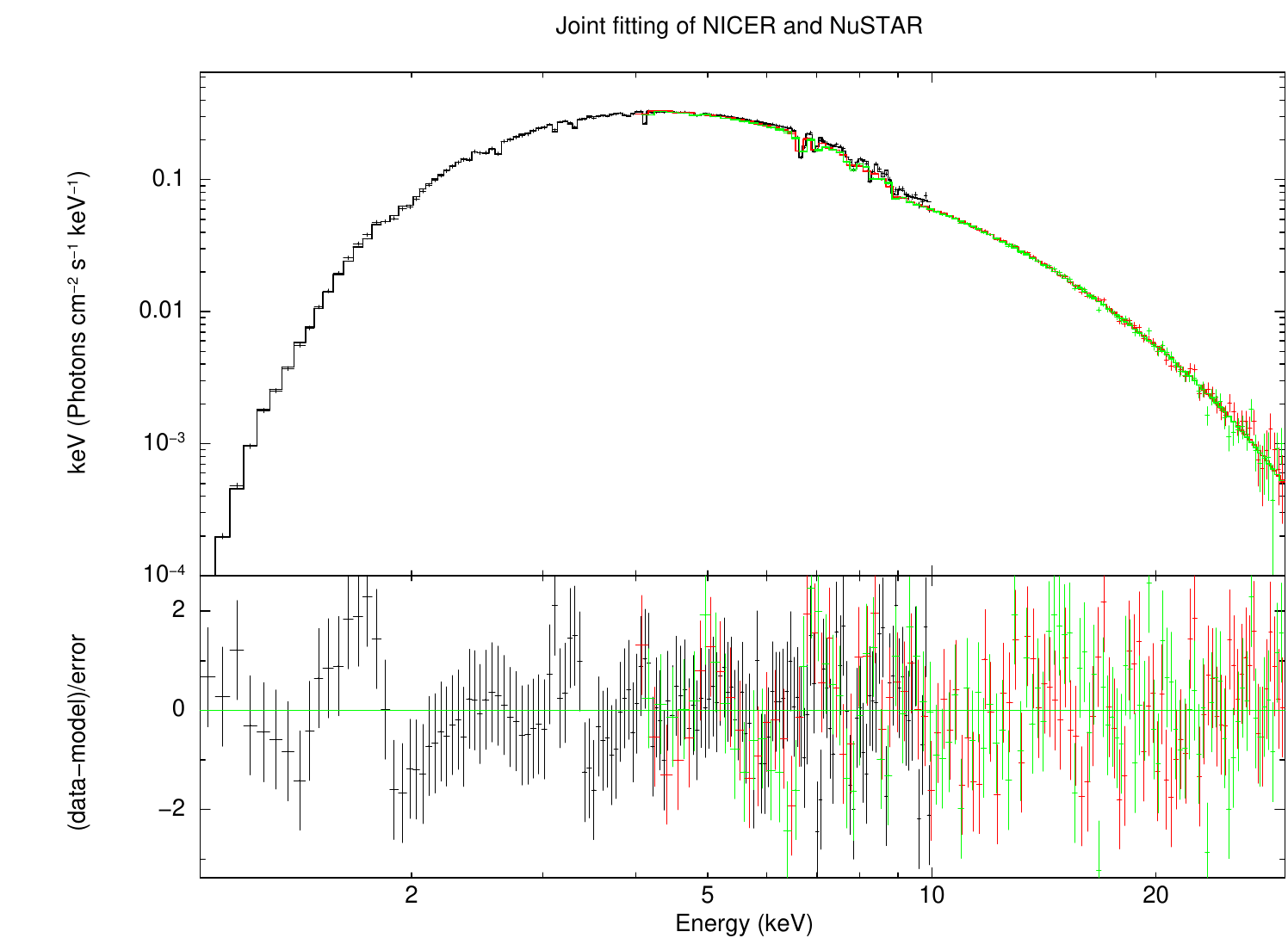}
    \caption{The left panel shows the joint fitting result by \texttt{TBabs$\times$edge$\times$thcomp$\times$diskbb} model. The right panel shows the result by \texttt{TBabs$\times$zxipcf$\times$edge$\times$edge$\times$thcomp$\times$diskbb} model }
    \label{nicer_nustar_fit0}
\end{figure*}

We first choose a thermally comptonized model \texttt{thcomp} times the multi-black body components \texttt{diskbb} from the accretion disk, and then times the Tuebingen-Boulder ISM absorption model \texttt{TBabs} with abundances as in (\citealp{Wilms2000}), and equivalent hydrogen column density $n_{\rm H}$ as a free parameter.
Cross-calibration between NICER and NuSTAR is carried out by the model \texttt{Crabcorr}(\texttt{plabs}) rather than a simple \texttt{constant} (\citealp{Steiner2010ApJ...718L.117S, Wang2020ApJ...899...44W}), which could multiply each spectrum by a power-law, applying corrections to both the slope via $\Delta\Gamma$ and normalization.
In this way, the responses of different detectors are cross-calibrated to return the same normalizations and power-law slopes for the Crab.
The model \texttt{plabs$\times$TBabs$\times$thcomp$\times$diskbb} gives a reduced $\chi^2=6441.94/350$. 
The best fit exhibits substantial residuals in $2-5$ keV, and it is not possible to derive reasonable disk parameters.
We should be mindful that during the outburst phase, there could be substantial additional absorption from neutral matters surrounding the central source and is consistently employed when modeling the spectrum during the obscured state of GRS~1915+105(\citealp{Neilsen2020ApJ...902..152N, Kong2021ApJ...906L...2K, Liu2022ApJ...933..122L}).  
To fit this kind of absorption, we incorporate an extra component in the form of a partial covering fraction absorption model, denoted as \texttt{pcfabs}. 
The residual also leaves a prominent broad dip structure around $\sim9$ keV without a known origin. 
We then use a \texttt{edge} model to fit this dip. In this way, the residual contains a structure between $1-3$ keV, leaving only narrow absorption lines. We note that the structures between $1-3$ keV are commonly seen in NICER's residuals, which can be attributed to calibration inaccuracy (e.g., the aluminum K edge/fluorescence at 1.56 keV, the silicon K edge from the detector window at 1.84 keV, and the gold M edge from XRC reflector gold coating at 2.2 keV\footnote{\url{https://heasarc.gsfc.nasa.gov/docs/nicer/data_analysis/workshops/NICER-CalStatus-Markwardt-2021.pdf}}).
The Fe \Rmnum{25} K-$\alpha$ line at $\sim6.7$ keV, the Fe \Rmnum{26} Ly-$\alpha$ at $\sim6.97$ keV, the Fe \Rmnum{25} K-$\beta$ line at $\sim7.8$ keV, and the Fe \Rmnum{26} Ly-$\beta$ line at $\sim8.27$ keV are very significant (see the left panel of Figure~\ref{nicer_nustar_fit0}).
To get a more physical insight for multiple absorption lines generation, we use a partial covering absorption by partially ionized material model \texttt{zxipcf}. 
This model uses a grid of XSTAR photoionized absorption models for the absorption, then assumes that this only covers some fraction of the source, while the remaining (1-f) of the spectrum is seen directly (\citealp{Miller2006A&A...453L..13M, Reeves2008MNRAS.385L.108R}). 
In this situation, we choose to keep the covering fraction (represented as 'f') fixed at 1 within the \texttt{zxipcf} model during the fitting process due to its degeneracy with other parameters and because it primarily falls within the range of $0.8-1$.
After the fitting process, we can still find a narrow dip structure at $\sim8$ keV, which can be fitted by another \texttt{edge} model. 
The total model is \texttt{TBabs$\times$zxipcf$\times$edge$\times$edge$\times$thcomp$\times$diskbb} (Model 1).
The results of fitting and parameters are shown in the right panel of Figure~\ref{nicer_nustar_fit0} and Table~\ref{spectral_fitting2}. 
We observe that the presence of the two additional dip structures around 7.6 keV and 8.8 keV can be attributed to the K-edge of Co at 7.7 keV and the K-edge of Cu at 8.97 keV or potentially due to the limitations of lower energy spectrum resolution within the 7-10 keV range, which may not accurately resolve the absorption lines.

From the fitting, the equivalent hydrogen column $n_{\rm H}$ from the interstellar medium (ISM), and possibly the cold matters surrounding the source, is $\sim5\times10^{22}$ cm$^{-2}$, which is consistent with the typical value during the persistent outburst (\citealp{Zoghbi2016, Kong2021ApJ...906L...2K, Liu2022ApJ...933..122L}).
The column density of \texttt{pcfabs} is $\sim7.5$ with a cover factor $0.46$. 
The disk temperature $kT_{\rm in}$ is 1.8 keV with the normalization 30.4. The \texttt{thcomp} has the $\Gamma=1.3$, $kT_{\rm e}=2.8$, and cover factor is 0.05.
With this model, although we can obtain acceptable residuals, we have to consider the reasonableness of the disk parameters.
By considering the inclination angle is 66$^{\circ}$ and a distance of 9 kpc, we can calculate the inner radius of the accretion disk $R_{\rm in}=\sqrt{N_{\rm disk}/\cos{\theta}}\ \xi\ f_{\rm col}^2\ D_{10}=9.5$ km (\citealp{Kubota1998PASJ...50..667K}), where $N_{\rm disk}$ is the \texttt{diskbb} normalization, $\theta$ is the inclination angle, $D_{10}$ is the distance in units of 10 kpc, $\xi=0.412$ is the general relativity correction factor, and $f_{\rm col}=1.7$ is the color-correction factor with the ``canonical'' value from \cite{Shimura1995ApJ...445..780S}.
Such a small value of the inner radius is far less and unreasonable for a Kerr black hole with 12 $M_{\odot}$ and a=0.98 ($R_{\rm ISCO}\sim1.6$ GM/c$^2$=28 km) (\citealp{Reid2014ApJ...796....2R}). 
If we assume that $R_{\rm in}=R_{\rm ISCO}$, we need $f_{\rm col}=2.9$, which is far larger than the value of a standard disk. 
Such a large $f_{\rm col}$ is only possible when the disk emission is relatively less dominant during the outburst of the Low-mass black hole binaries (\citealp{Merloni2000MNRAS.313..193M, Ren2022ApJ...932...66R}).

In this situation, the nature of the disk itself may be far beyond the geometric thin disk, or the disk may be so encased in the warm layer (\citealp{Zhang2000Sci...287.1239Z, Gronkiewicz2020A&A...633A..35G}) that the radiation from the disk is completely diluted. Such kind of warm layer is stronger in the case of the higher accretion rate and the greater magnetic field strength (\citealp{Gronkiewicz2020A&A...633A..35G}). 
Following \cite{Zhang2000Sci...287.1239Z, Pszota2007ApJ...663.1201P}, we replace \texttt{diskbb} to \texttt{comptt} to model the disk emission with saturated Compton scattering (\citealp{Titarchuk1994ApJ...434..570T}). And we found the \texttt{pcfabs} is not needed during the spectral fitting. 
The total model is \texttt{TBabs$\times$zxipcf$\times$edge$\times$edge$\times$thcomp$\times$comptt} (Model 2). The parameters are shown in Table~\ref{spectral_fitting2}. 
The $T_0$ value for the \texttt{comptt} model is 0.23 keV, and it is linked to $kT/2.7f_{\rm col}$ during the fitting, where $f_{\rm col}=1.7$, representing a standard disk (refer to \citealp{Pszota2007ApJ...663.1201P} for more details).
At this time, we found the optical depth $\tau$ is up to 57.
This implies that the total emission from the accretion disk undergoes saturated Compton scattering in some low-temperature, optically thick gas before the seed photons are up-scattered to higher energy levels. Because the optical depth is large, the \texttt{comptt} is equivalent to a 1 keV black body. 
Also, since $f_{\rm cov}$ is close to 1, the model \texttt{thcomp$\times$comptt} can be simplified to the \texttt{nthComp} model with \texttt{inp\_type}=0 indicating blackbody seed photons.

Based on the analysis mentioned above, we are inclined to use the 
\texttt{TBabs$\times$zxipcf$\times$edge$\times$edge$\times$nthComp} (Model 3) for further spectral analysis.
The fitting results are shown in Table~\ref{spectral_fitting2}.
The overall continuum spectrum can be fitted with an \texttt{nthComp} model with $\Gamma=3.3$, $T_{\rm bb}=1.07$, and $KT_{\rm e}=3.9$. 

\begin{figure*}
    \centering\includegraphics[width=0.49\textwidth]{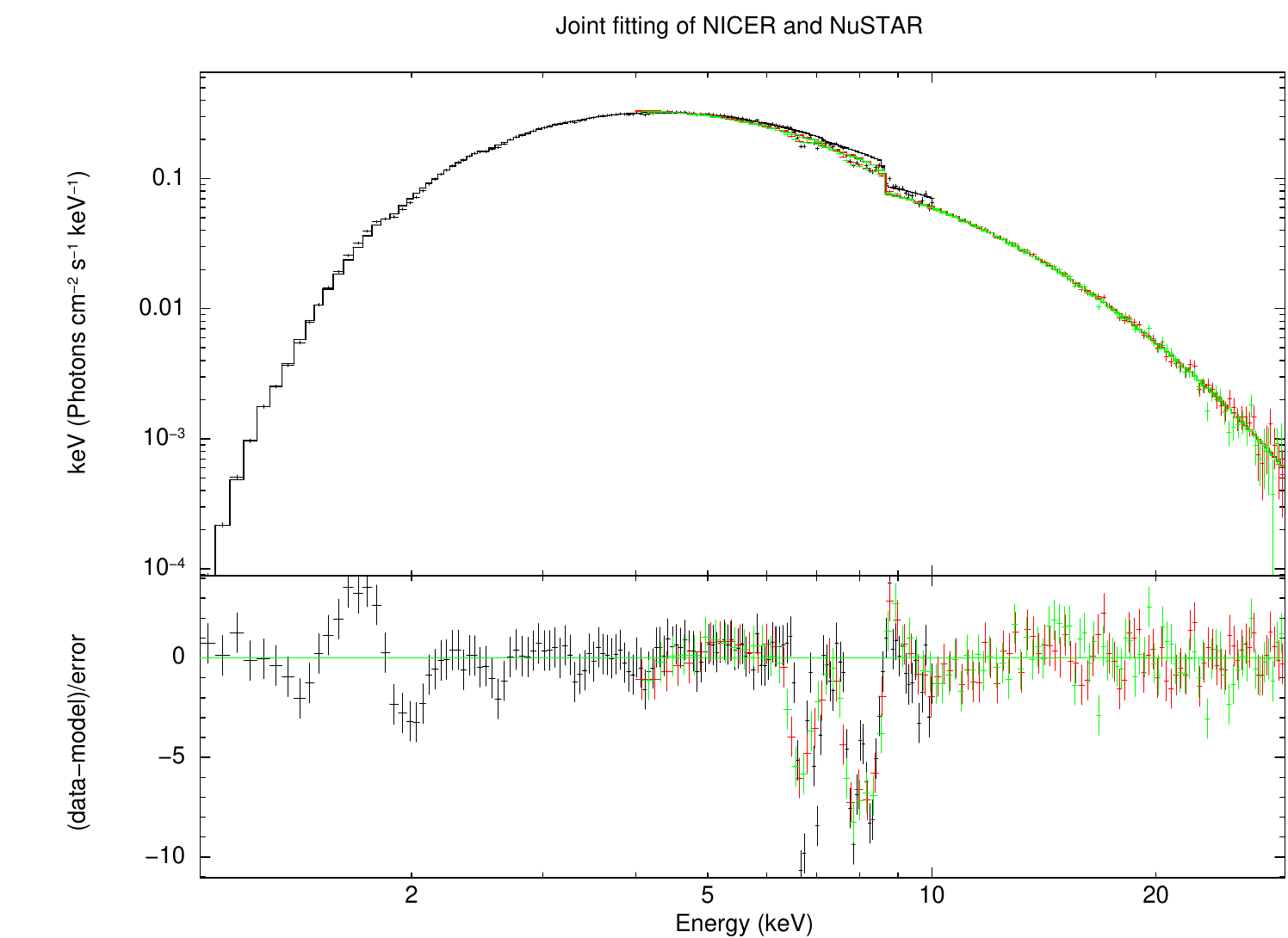}
    \centering\includegraphics[width=0.49\textwidth]{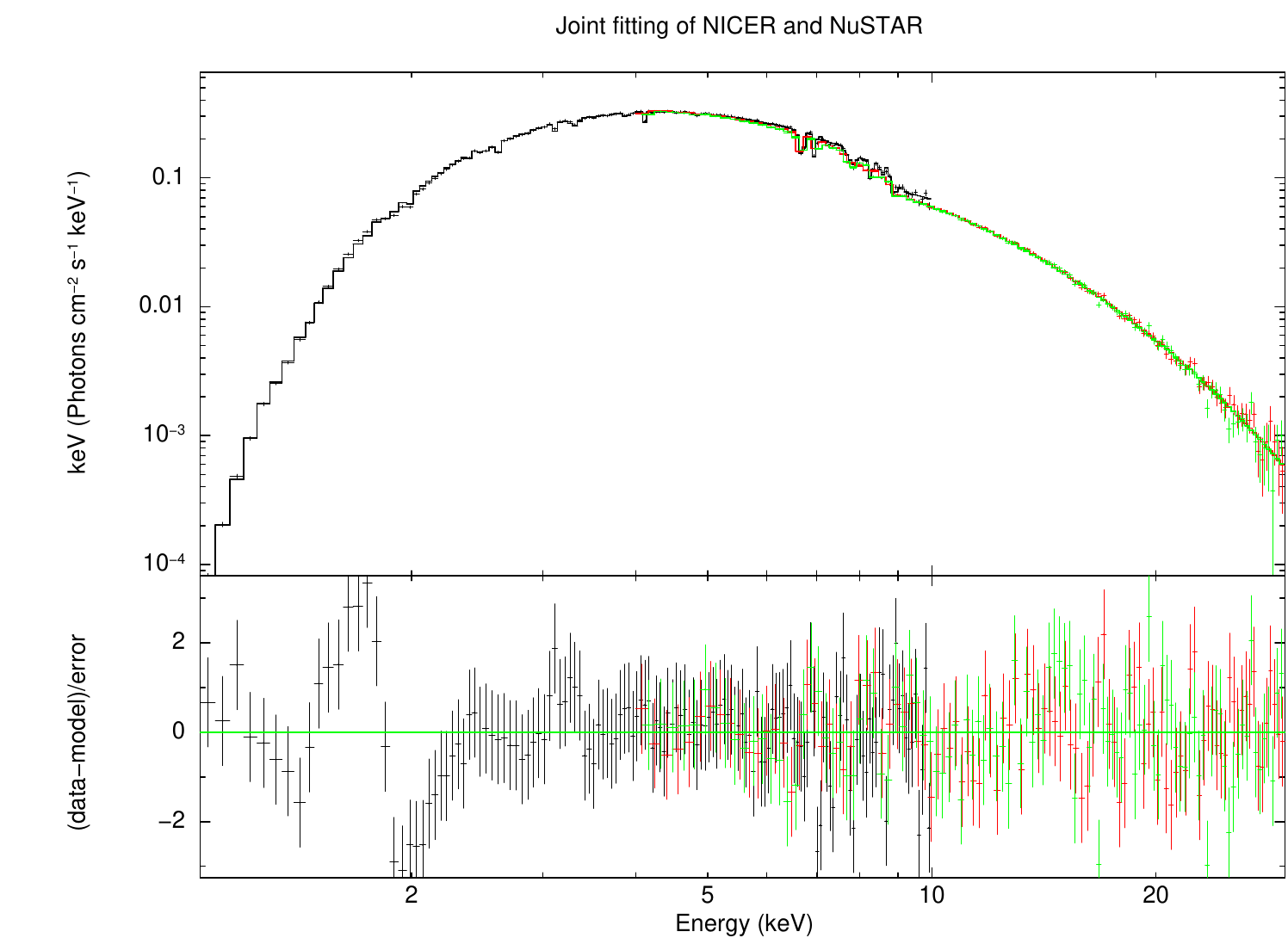}
    \caption{The left panel shows the joint fitting result by \texttt{TBabs$\times$edge$\times$nthComp} model. The right panel shows the result by \texttt{TBabs$\times$zxipcf$\times$edge$\times$edge$\times$nthComp} model }
    \label{nicer_nustar_fit}
\end{figure*}

\begin{table*}
\caption{Spectral parameters of joint fitting of NICER+NuSTAR}
\centering
\begin{tabular}{ccccc}
\hline
\hline
Component & Parameters & Model 1 & Model 2 & Model 3 
\\
\hline
TBabs & $n_{\rm H}\ (10^{22}\ \rm cm^{-2})$ & $5.7_{-0.1}^{-0.1}$ & $4.9_{-0.1}^{+0.1}$ & $5.2_{-0.1}^{+0.1}$ 
\\
\hline
zxipcf & $N_{\rm h}\ (10^{22}\ \rm cm^{-2})$ & $81_{-16}^{+19}$ & $41_{-11}^{+7}$ & $47_{-11}^{+29}$ 
\\
& log$\ \xi$ & $4.1_{-0.1}^{+0.1}$ & $4.0_{-0.1}^{+0.1}$ & $4.1_{-0.1}^{+0.1}$
\\
& f & 1 (fixed) & 1 (fixed) & 1 (fixed)
\\
& Redshift & 0 (fixed) & 0 (fixed) & 0 (fixed)
\\
\hline
pcfabs & $n_{\rm H}\ (10^{22}\ \rm cm^{-2})$ & $7.5_{-1.1}^{+0.6}$ & ... & ...
\\
& $f_{\rm cov}$ & $0.46_{-0.02}^{+0.04}$ & ... & ...
\\
\hline
edge1 & $E_{\rm edge}1$ (keV) & $7.62_{-0.03}^{+0.03}$ & $7.59_{-0.01}^{+0.04}$ & $7.60_{-0.04}^{+0.05}$ 
\\
& $\tau_{\rm edge}1$ & $0.09_{-0.01}^{+0.01}$ & $0.11_{-0.01}^{+0.01}$ &$0.12_{-0.02}^{+0.01}$ 
\\
\hline
edge2 & $E_{\rm edge}2$ (keV) & $8.75_{-0.04}^{+0.03}$ & $8.76_{-0.01}^{+0.02}$ &$8.76_{-0.02}^{+0.02}$ 
\\
& $\tau_{\rm edge}2$ & $0.19_{-0.04}^{+0.04}$ & $0.25_{-0.01}^{+0.02}$ &$0.25_{-0.03}^{+0.02}$ 
\\
\hline
nthComp & $\Gamma$ & ... & ... & $3.3_{-0.1}^{+0.1}$ 
\\
& $kT_e$ (keV) & ... & ... & $3.9_{-0.8}^{+0.4}$ 
\\
& $T_{\rm bb}$ (keV) & ... & ... & $1.07_{-0.01}^{+0.01}$ 
\\
& norm & ... & ... & $0.15_{-0.01}^{+0.01}$ 
\\
\hline
thcomp & $\Gamma$ & $1.3_{-0.1}^{+0.1}$ & $3.2_{-0.1}^{+0.1}$ & ... 
\\
& $kT_e$ (keV) & $2.8_{-0.1}^{+0.1}$ & $3.8_{-0.1}^{+0.2}$ & ... 
\\
& $f_{\rm cov}$ & $0.05_{-0.01}^{+0.01}$ & $0.9_{-0.2}^{+0.1}$ & ... 
\\
\hline
diskbb & $kT_{\rm in}$ (keV) & $1.77_{-0.02}^{+0.03}$ & ... & ... 
\\
& norm & $30.4_{-2.2}^{+1.9}$ & ... & ... 
\\
\hline
comptt & $T_0$ (keV) & ... & $0.23^{a}$ & ...
\\
& $kT$ (keV) & ... & $1.06_{-0.01}^{+0.01}$ & ...
\\
& $\tau$ & ... & $57_{-3}^{+2}$ & ...
\\
& norm & ... & $0.74_{-0.03}^{+0.02}$ & ...
\\
\hline
Crabcorr & $\Delta \Gamma_{\rm FPMA}$ & $0.15_{-0.02}^{+0.02}$ & $0.15_{-0.01}^{+0.02}$ & $0.16_{-0.03}^{+0.02}$ 
\\
(plabs) & $K_{\rm FPMA}$ & $1.3_{-0.1}^{+0.1}$ & $1.3_{-0.1}^{+0.1}$ & $1.3_{-0.1}^{+0.1}$ 
\\
& $\Delta \Gamma_{\rm FPMB}$ & $0.14_{-0.02}^{+0.02}$ & $0.15_{-0.01}^{+0.02}$ & $0.16_{-0.02}^{+0.02}$ 
\\
& $K_{\rm FPMB}$ & $1.2_{-0.1}^{+0.1}$ & $1.3_{-0.1}^{+0.1}$ & $1.3_{-0.1}^{+0.1}$ 
\\
\hline
Fitting & $\chi_{\rm red}^{2}/d.o.f$ & 1.1/339 & 1.1/340 & 1.1/341 
\\
\hline
unabsorbed $L_{\rm x}$  & ergs/s & $5.25\times10^{37}$ & $3.77\times10^{37}$ & $3.89\times10^{37}$ 
\\
($1-10$ keV) 
\\
\hline
unabsorbed $L_{\rm x}$  & ergs/s & $5.53\times10^{36}$ & $5.37\times10^{36}$ & $5.45\times10^{36}$ 
\\
($10-25$ keV) 
\\
\hline
\hline
    \end{tabular}
    \label{spectral_fitting2}
\begin{list}{}{}
    \item[Note]{: Uncertainties are reported at the 90\% confidence interval and were computed using MCMC (Markov Chain Monte Carlo) of length 10,000.\\
    $a$: $T_0$ is linked to $kT/2.7f_{\rm col}$, where $f_{\rm col}=1.7$.}
\end{list}
\end{table*}

\subsubsection{Spectral evolution}

\begin{figure*}
    \centering\includegraphics[width=0.49\textwidth]{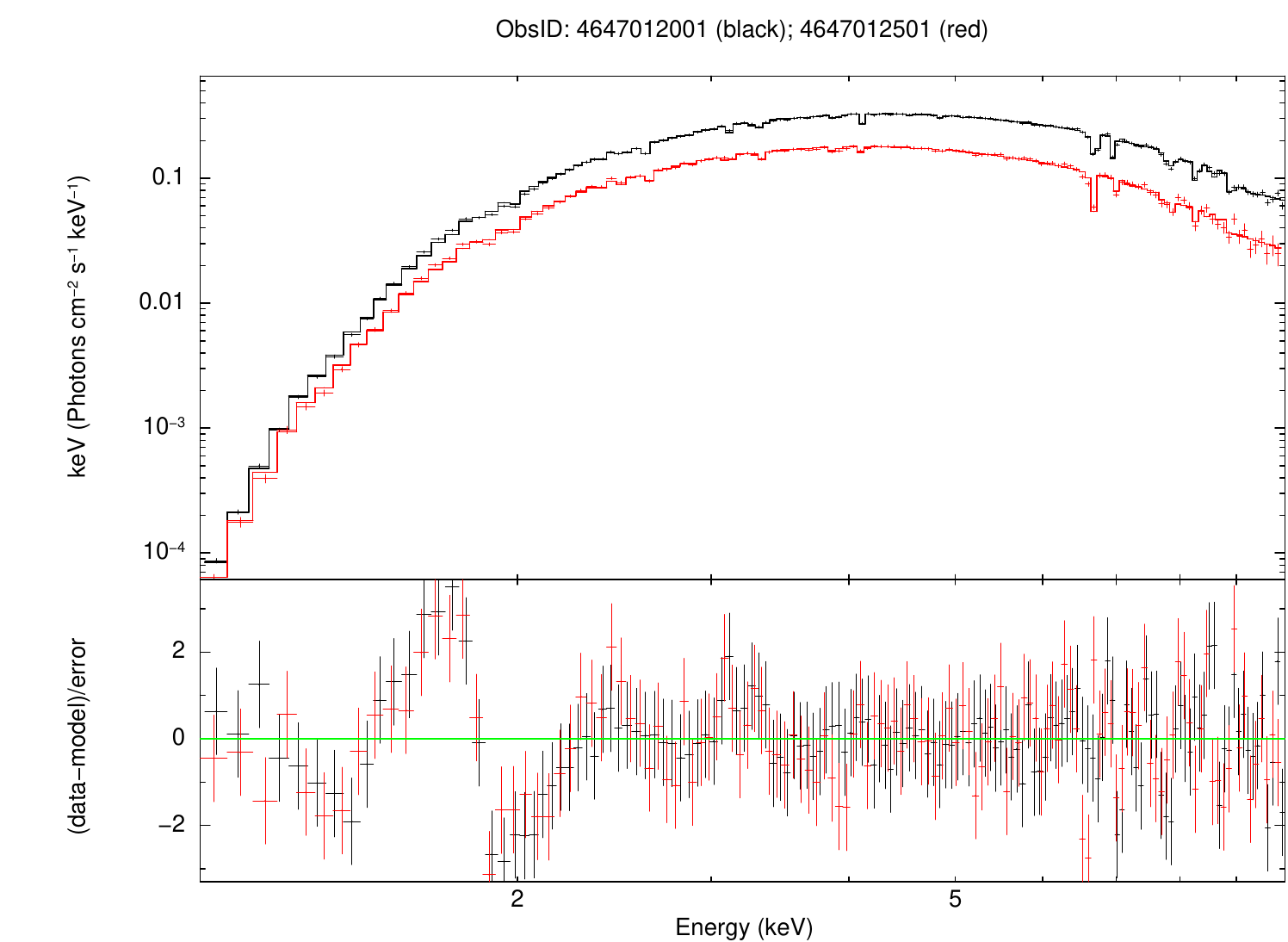}
    \centering\includegraphics[width=0.49\textwidth]{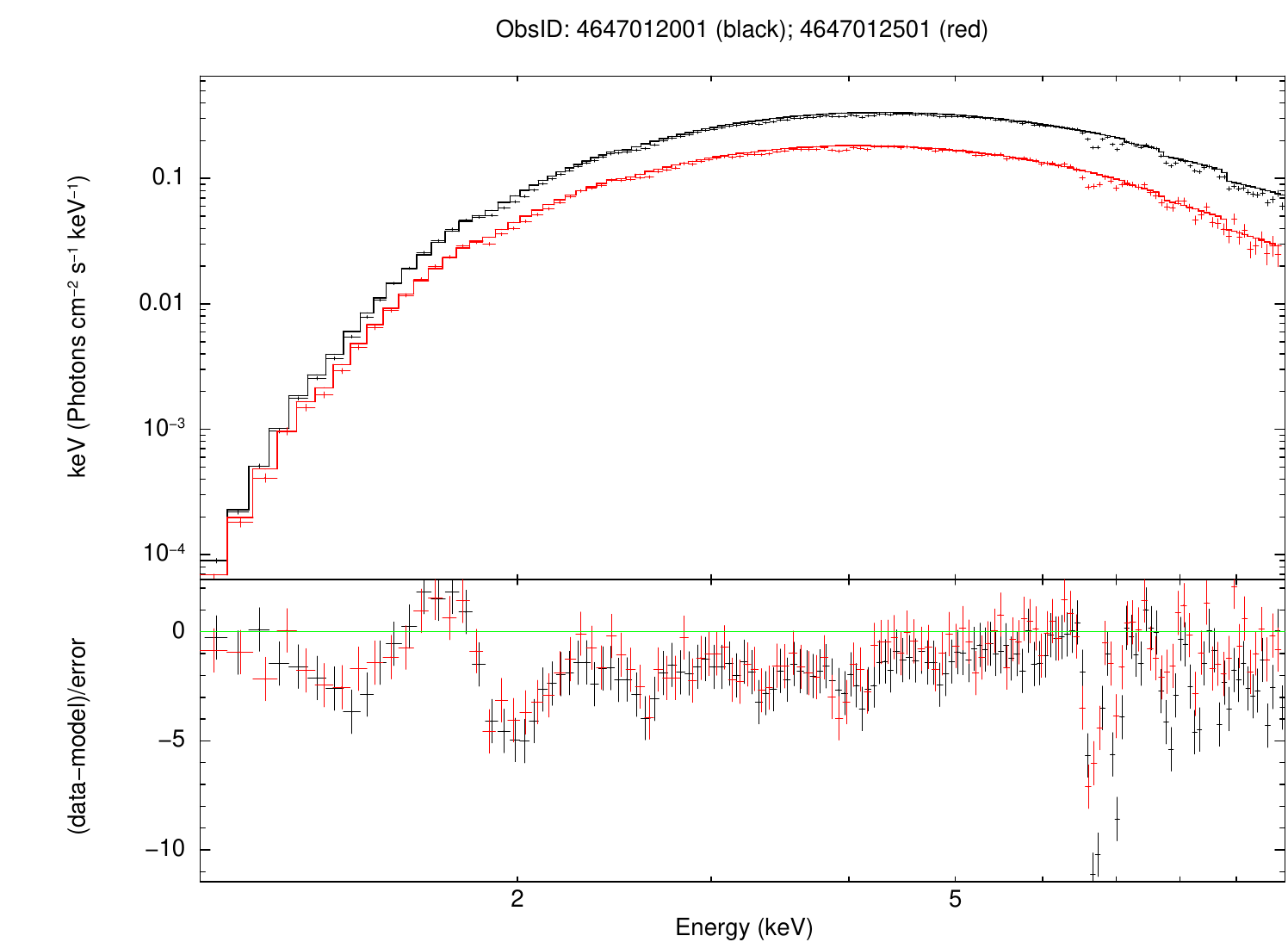}
    \caption{The left panel displays the residuals following the inclusion of the \texttt{zxipcf} model of ObsID: 4647012001 (black points) and ObsID: 4647012501 (red points). The right panel shows the residuals excluding the \texttt{zxipcf} model.}
    \label{nicer_fit}
\end{figure*}

We have performed an analysis of the spectral evolution using all available NICER observations and by employing the \texttt{TBabs$\times$zxipcf$\times$edge$\times$edge$\times$nthComp} model (Model 3). 
The observations used for spectral evolution analysis are identical to the data utilized for timing analysis.
Throughout this analysis, we maintain a fixed electron temperature of 4 keV as determined through the joint (broader band) modeling discussed in the previous section.
As representative examples for spectral fitting, we choose ObsID 4647012001, which exhibits a QPO in Epoch 1, and ObsID 4647012501, which displays peaked noise (or a broad QPO with a Q factor of approximately 2) in Epoch 3.
The fitting results are shown in Figure~\ref{nicer_fit} and Table~\ref{spectral_fitting1}. 
For ObsID 464712001, the column density $N_{\rm h}$ is $\sim50\times10^{22}$ cm$^{-2}$ with a high ionization degree log $\xi\ \sim4.1$.
The property of the absorption in ObsID 4647012001 changes into lower column density $\sim13\times10^{22}$ cm$^{-2}$ and ionization degree log $\xi\ \sim3.4$.

\begin{figure}
    \resizebox{\hsize}{!}{\includegraphics{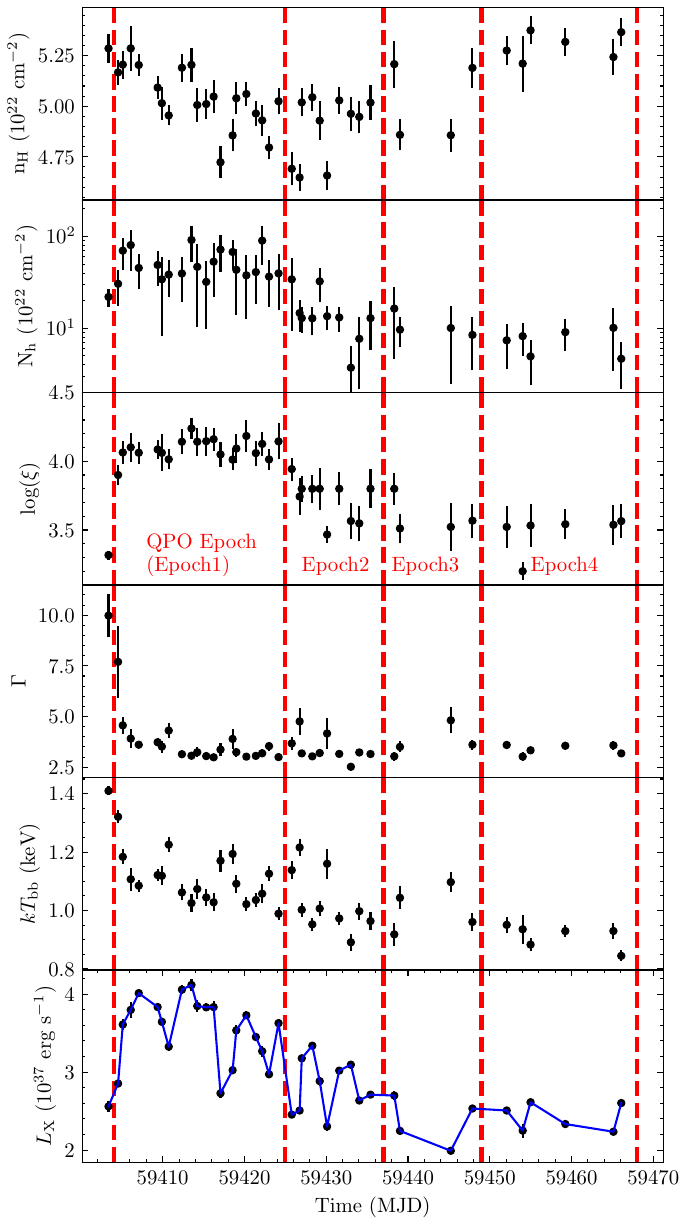}}
    \caption{The figure presents the fitting outcomes derived from the \texttt{TBabs$\times$zxipcf$\times$edge$\times$edge$\times$nthComp} model (Model 3). Illustrating the evolution of neutral hydrogen absorption $n_{\rm H}$ from the interstellar medium (ISM) and cold dust. It further details the wind column density $N_{\rm h}$, the degree of wind ionization represented by log$(\xi)$, the photon index $\Gamma$, the seed photon temperature $kT_{\rm bb}$, and the luminosity before absorption.  Within the \texttt{zxipcf} component, the covering fraction and redshift are held constant at values of 1 and 0, respectively. From the joint fitting results, the electron temperature $kT_{\rm e}$ in the \texttt{nthComp} model remains fixed at 4 keV.}
    \label{spec_evo}
\end{figure}

During the fitting process, we noticed that the parameters of the two edge models remained largely unchanged for most of the observations. 
Therefore, we fixed $E_{\rm edge1}$ at 7.6 keV, $\tau_{\rm edge1}$ at 0.1, $E_{\rm edge2}$ at 8.8 keV and $\tau_{\rm edge2}$ at 0.18 for two \texttt{edge} models.
By multiplying a \texttt{cflux} on \texttt{nthComp}, we can calculate the unabsorbed flux in $1-10$ keV of the source.
The results of the evolution of the continuum spectrum parameters can be found in Figure~\ref{spec_evo}.
The neutral hydrogen column density $n_{\rm H}$ in \texttt{TBabs} remains relatively constant, predominantly within the range of $4.76\times10^{22}$ to $5.25\times10^{22}$ cm$^{-2}$, with no significant evolution.
For the ionized absorption component that we are most concerned about, we find that the column density $N_{\rm h}$ and the ionization degree log $\xi$ exhibits higher values $\sim50\times10^{22}$ cm$^{-2}$ and $\sim4.1$ in QPO Epoch (Epoch 1), and then decrease into lower values in Epoch 2, Epoch 3 and Epoch 4.
The photon index $\Gamma$ decreases from 10 to 2.5 during the whole re-brightening. 
The primary decline occurs during the first week after the re-brightening begins, and after that, the value remains stable at 2.5. 
For a given $kT_{\rm e}=4$ and photon index $2.5<\Gamma<10$, we can estimate the optical depth $1<\tau<7$ by the equation A1 in \cite{Zdziarski1996MNRAS.283..193Z}.
The blackbody temperature, $kT_{\rm bb}$, decreases from 1.4 to 0.8. 
During the re-brightening, $L_{\rm X}$ has plenty of dips that correspond to spikes of $\Gamma$ and $kT_{\rm bb}$ at the same time. The total luminosity $L_{\rm X}$ can be calculated by $4\pi D^2F_{\rm X}$, where $D=9$ kpc.

Here, we must emphasize that it is possible that the results on the evolution of photon index $\Gamma$ and $kT_{\rm e}$ are based only on phenomenology and not on real results.
Due to the limitations of the NICER energy band, we can only approximate the value of kTe based on the only NuSTAR observation. 
The evolution of $\Gamma$ obtained by fixing $kT_{\rm e}$ may be incorrect since $kT_{\rm e}$ itself is likely to degeneracybe evolving during the re-brightening. 
Since the parameters of the ionized absorption model \texttt{zxipcf} do not have degeneracy with $\Gamma$ and $kT_{\rm e}$, we consider their evolution to be reliable.

\begin{table*}
\caption{Spectral parameters of joint fitting of NICER}
\centering
\begin{tabular}{cccccc}
\hline
\hline
Component & Parameters & 4647012001 & 4647012001 & 4647012501 & 4647012501
\\
\hline
TBabs & $n_{\rm H}\ (10^{22}\ \rm cm^{-2})$ & $5.1_{-0.1}^{+0.1}$ & $5.1_{-0.1}^{+0.1}$ & $4.9_{-0.1}^{+0.1}$ & $4.9_{-0.1}^{+0.1}$
\\
\hline
zxipcf & $N_{\rm h}\ (10^{22}\ \rm cm^{-2})$ & $51_{-17}^{+47}$ & $49_{-13}^{+26}$ & $10_{-3}^{+4}$ & $10_{-2}^{+11}$
\\
& log$\ \xi$ & $4.1_{-0.1}^{+0.1}$ & $4.1_{-0.1}^{+0.1}$ & $3.5_{-0.1}^{+0.1}$ & $3.5_{-0.1}^{+0.3}$
\\
& f & 1 (fixed) & 1 (fixed) & 1 (fixed) & 1 (fixed)
\\
& Redshift & 0 (fixed) & 0 (fixed) & 0 (fixed) & 0 (fixed)
\\
\hline
edge1 & $E_{\rm edge}1$ (keV) & $7.6_{-0.1}^{+0.1}$ & 7.7 (fixed) & $7.5_{-0.1}^{+0.2}$ & 7.7 (fixed)
\\
& $\tau_{\rm edge}1$ & $0.10_{-0.02}^{+0.02}$ & 0.1 (fixed) & $0.09_{-0.06}^{+0.06}$ & 0.1 (fixed)
\\
\hline
edge2 & $E_{\rm edge}2$ (keV) & $8.8_{-0.1}^{+0.1}$ & 8.8 (fixed) & $8.6_{-0.1}^{+0.1}$ & 8.8 (fixed)
\\
& $\tau_{\rm edge}2$ & $0.18_{-0.04}^{+0.04}$ & 0.18 (fixed) & $0.23_{-0.1}^{+0.1}$ & 0.18 (fixed)
\\
\hline
nthComp & $\Gamma$ & $3.7_{-0.3}^{+0.2}$ & $3.7_{-0.2}^{+0.2}$ & $4.5_{-0.5}^{+1.0}$ & $4.8_{-0.5}^{+0.8}$
\\
& $kT_e$ (keV) & 4 (fixed) & 4 (fixed) & 4 (fixed) & 4 (fixed) 
\\
& $T_{\rm bb}$ (keV) & $1.12_{-0.03}^{+0.02}$ & $1.12_{-0.02}^{+0.02}$ & $1.08_{-0.03}^{+0.04}$ & $1.10_{-0.04}^{+0.03}$
\\
& norm & $0.14_{-0.01}^{+0.01}$ & $0.14_{-0.01}^{+0.01}$ & $0.08_{-0.01}^{+0.01}$ & $0.08_{-0.01}^{+0.01}$
\\
\hline
unabsorbed Luminosity ($1-10$ keV) & ergs/s & $3.84\times10^{37}$ & $3.83\times10^{37}$ & $2.01\times10^{37}$ & $1.99\times10^{37}$
\\
\hline
Fitting (System Error 1\%) & $\chi_{\rm red}^{2}/d.o.f$ & 1.13/135 & 1.12/139 & 1.25/119 & 1.28/119
\\
\hline
\hline
    \end{tabular}
    \label{spectral_fitting1}
\begin{list}{}{}
    \item[Note]{: Uncertainties are reported at the 90\% confidence interval and were computed using MCMC (Markov Chain Monte Carlo) of length 10,000.}
\end{list}
\end{table*}

\section{Discussions}

This work presents a detailed timing-spectral analysis of the NICER's observations relative to the X-ray re-brightening of GRS~1915+105 in 2021. Compared to the hard state before entering the low flux state (t $<$ MJD 58600), although the flux below 20 keV of the whole re-brightening is similar to the hard state, the flux above 20 keV has significantly decreased (see Figure~\ref{maxi-bat}). The key finding of our analysis is that the spectral and timing properties of the source during the re-brightening phase exhibit significant differences compared to those typical of the soft and hard states of black hole binaries. 

The emission is dominated by the Comptonization of seed photons within the optically thick, low-temperature plasma and displays ionized absorption lines. The numerous absorption lines in the spectrum indicate that photons have interacted with an ionized absorbing medium, whose properties undergo significant changes throughout the re-brightening phase. More specifically the plasma evolves from higher column density $N_{\rm h}\sim50\times10^{22}$ cm$^{-2}$ and ionization degree log $\xi\sim4$ to the lower column density $N_{\rm h}\sim10\times10^{22}$ cm$^{-2}$ and ionisation degree log $\xi\sim3.5$. 
What's even more intriguing is that during this soft X-ray re-brightening, QPOs are observed up to an energy of 18 keV (see the upper panel of Figure~\ref{qpo_rms_E}). The QPOs exhibit a relatively weak evolution in terms of frequency spanning the range of $0.17-0.21$ Hz.
By comparing Figure~\ref{Lor_evo} and Figure~\ref{spec_evo}, it becomes evident that the QPOs only manifest within the initial 20 days following the onset of the re-brightening. This coincides with the phase in which the absorbing medium exhibits high column density and ionization degree.
After the PDS evolves from a QPO signal superimposed on a low-frequency broadband-limited noise into a high-frequency peaked noise superimposed on a low-frequency broadband noise (see Figure~\ref{PDS_fitting}), the column density and ionization degree decrease. 
Figure~\ref{qpo_evo} clearly shows the QPO's properties.
In particular, although the frequency of QPO increases with time from 0.17 Hz to 0.21 Hz, there are three sudden decreases, which correspond precisely to the sudden increase in the $\Gamma$, and $kT_{\rm bb}$ parameters and the decrease in flux at MJD 58411, MJD 58417, and MJD 58423 (see Figure~\ref{spec_evo}).
As mentioned earlier, in using the \texttt{nthComp} model, we assume a constant electron temperature obtained from the only NICER and NuSTAR joint observation during the re-brightening episode. 
In addition, the model we use differs significantly from typical models employed in modeling the X-ray emission of black hole binary systems. 
We can get a good fitting by using a canonical model like \texttt{tbabs$\times$pcfabs$\times$thcomp$\times$diskbb}, but as we discussed above, this model can not give a reasonable inner radius of the accretion disk. The \texttt{diskbb} should be abandoned in the fitting; this means that the emission from the accretion disk might be totally covered by the optically thick and low-temperature electrons, which can be modeled by a saturated Compton scattering model \texttt{comptt}.
In summary, our findings regarding the soft X-ray re-brightening in 2021 indicate that the radiation is predominantly dominated by Compton-thick low-temperature electrons accompanied by a QPO signal below 18 keV. Furthermore, the appearance and disappearance of these QPOs are correlated with the characteristics of the absorbing material surrounding the source. Specifically, QPOs only manifest when the absorbing medium shows high column densities and a high ionization degree.
In our following discussion, the purpose is to discuss the origin of such unique low-frequency QPOs based on the properties of QPOs and the results of the spectral fitting.

\subsection{The specificity of state and QPO property}

As we know, both in typical transient BHs or GRS~1915+105, the spectral states can be classified as soft states (SS), hard states (HS), and intermediate states (IMS) (\citealp{Belloni2010LNP...794...53B}). The LFQPOs are usually found in the hard state or intermediate states (\citealp{Ingram2019NewAR..8501524I}).

The peak unabsorbed/intrinsic luminosity in $1-10$ keV of this re-brightening is $\sim4\times10^{37}$ erg s$^{-1}$ which is 0.03 $L_{\rm Edd}$ (assuming a BH mass of 12 $M_{\odot}$ (\citealp{Reid2014ApJ...796....2R}).
This flux level is comparable to that in the decay at MJD $58220-58330$ and to the peak flux of the soft X-ray re-brightening in 2020 (MJD 59100) (\citealp{Koljonen2021A&A...647A.173K, Motta2021MNRAS.503..152M}).
However, a significant difference is that the decay state is more consistent with a hard state.
Compared with the soft state around MJD 58000 (see \citealp{Motta2021MNRAS.503..152M}), the flux of the soft X-ray re-brightening in 2021 is lower, and the duration of the soft emission is shorter. 

Meanwhile, the type-C QPOs around 0.17 Hz emerge during the initial phase of the soft X-ray re-brightening in 2021.
Such time-varying signals are typically more prevalent in the hard state as opposed to the soft state. This re-brightening phenomenon, once again, deviates from the characteristics of a typical soft state in this regard.
\cite{Zhang2020MNRAS.494.1375Z} presented a systematic analysis of the type-C quasi-periodic oscillations (QPOs) in GRS 1915+105 using RXTE data. 
The statistical analysis reveals that the frequency distribution of type-C QPOs falls within the range of 0.4 to 6.4 Hz, with the upper limit extending to 40 keV in terms of high-energy observations.
The case is different in our work. We have observed the QPOs in the initial phase of the re-brightening fall within a significantly lower frequency range of $\sim0.17-0.21$\, Hz. 
As shown in the upper panel of Figure~\ref{qpo_rms_E}, although the increasing trend of the QPO RMS spectrum is similar to that of the QPO in \cite{Zhang2020MNRAS.494.1375Z}, the maximum energy that QPO and the frequency of QPO can reach are entirely different.
Such a trend of increasing RMS with multiple scattering events indicates that seed photons undergoing more scatterings exhibit higher RMS values. The QPO signal doesn't originate from the incident photons; instead, it arises from the scattering plasma, that is from the electrons. This conclusion aligns with the origin of QPOs in typical black hole systems being consistent with the corona.
Compared to the corona temperatures reaching $50-100$ keV during typical outbursts, an electron temperature of 4 keV makes it challenging for seed photons to scatter to energies above 20 keV. Consequently, this prevents us from observing QPO signals at higher energy levels.

\subsection{The properties of the absorber}

Absorption lines with blue-shift speed from hundred km/s to 0.1c  provide evidence for the presence of disk winds driven by thermal pressure (\citealp{Tombesi2010A&A...521A..57T, Done2018MNRAS.473..838D}), radiation pressure (e.g., \citealp{Higginbottom2015ApJ...807..107H}), or magnetic pressure or centrifugal forces (\citealp{Fukumura2010ApJ...715..636F, Fukumura2017NatAs...1E..62F}). 
Based on the fitting results, we observe that blue-shift and red-shift values are nearly zero for nearly all observations, which is why we held them constant at 0. This suggests that the absorber exhibited slow movement along the line of sight.
In our case, it may be more in line with a ``failed wind'' during the obscured state of GRS~1915+105 with low flux level (\citealp{Miller2020ApJ...904...30M}).
From the parameters of the \texttt{zxipcf} model, we can estimate the upper limit of the launch radius $R_{\rm launch}$ according to the following equation (\citealp{Tarter1969}):
\begin{equation}
R_{\rm Launch} \leq L\times(N_{\rm h}\xi)^{-1},
\end{equation}
where the L is the luminosity.
Here, we take the results at the re-brightening peak, the L$\sim3.83\times10^{37}$ erg s$^{-1}$ (0.025 L$_{\rm edd}$), $N_{\rm h}\sim34-98\times10^{22}$ cm$^{-2}$, and log $\xi\sim4.1$, and get the $R_{\rm Launch}\leq3-9\times10^{4}$ km.  
\cite{Dubus2019} considered the contribution from radiation pressure at luminosity close to Eddington (\citealp{Proga2002}; \citealp{Done2018MNRAS.473..838D}). Accordingly, the equation for $T_{\rm IC}$ and $R_{\rm IC}$ is shown as follow:
\begin{equation}
\frac{T_{\rm IC}}{10^{7}\ \rm K}=\begin{cases}4.2-4.6\times \rm log(l/0.02)\ \rm if\ \emph{l} < 0.02 \\0.36\times(l/0.02)^{1/4}\ \rm if\ \emph{l} \ge 0.02\end{cases},
\end{equation}
\begin{equation}
R_{\rm IC}\approx10^{12}\times\frac{M}{10\ M_{\odot}}\times\frac{10^{7}\ \rm K}{T_{\rm IC}}\times(1-\sqrt{2}\times l),
\end{equation}
where $l=L/L_{\rm edd}$.
The Compton temperature $T_{\rm IC}$ and Compton radius $R_{\rm IC}$ can be estimated as $0.69\times10^{7}\ \rm K$ and $1.1\times10^{7}\ \rm km$, respectively.
Since the $R_{\rm Launch}$ is $0.003 R_{\rm IC}$, which is far less than $0.2 R_{\rm IC}$, the absorber is not likely thermal driven (\citealp{Woods1996ApJ...461..767W}).
Therefore, we assume the absorber is a Compton-thick magnetically driven failed disk wind.
Similarly, we can also estimate the $R_{\rm Launch}\leq4.5-9\times10^5$ km, $T_{\rm IC}\sim5.58\times10^7$ K, and $R_{\rm IC}\sim2.1\times10^{6}$ km at lower $N_{\rm h}\sim7-14\times10^{22}$ erg s$^{-1}$, log $\xi\sim3.5$ and Luminosity $L\sim2\times10^{37}$ erg s$^{-1}$ (0.01 $L_{\rm Edd}$).
The $R_{\rm Launch}\sim0.21-0.43\ R_{\rm IC}$ is near the threshold 0.2 $R_{\rm IC}$, which means that the thermally driven case can not be excluded here.

The above calculation makes it easy to find that the driving mechanism and position of the failed wind vary significantly at different Luminosity or accretion rates.
In the QPO epoch, the luminosity is high, and the wind is generated closer to the black hole ($R_{\rm Launch}\sim0.003 R_{\rm IC}$).
In this case, it may be driven by the magnetic pressure on the disk and maintained in equilibrium with gravity (although the redshift/blueshift is very small, we cannot exclude that the disk wind has a significant speed in the direction perpendicular to the disk because GRS~1915+105 is highly inclined binary. 
At lower luminosity, when there is no QPO signal, the winds are generated further away from the black hole ($R_{\rm Launch}\sim0.21-0.43\ R_{\rm IC}$) when they might be thermally driven.

\subsection{The origin of QPO}

We found a long-duration LFQPO with $\sim0.17-0.21$ Hz appearing during the soft X-ray re-brightening. 
Compared with the properties of typical type-C QPOs, the frequency is lower than the lowest value in \cite{Zhang2020MNRAS.494.1375Z}. 
The maximum energy of the observed QPO is 18 keV, which is much lower than the oscillation at 100 keV that a type-C QPO can achieve (\citealp{Huang2018ApJ...866..122H, Ma2021NatAs...5...94M}).
Compared to the latter one, the QPO is found to be produced in a Compton-thick low-temperature gas.
Determining the composition of the accretion disk from the energy spectrum is challenging due to the potential complete shielding by the surrounding Compton-thick gas. This situation makes it impossible to collect information about the radiation within the Compton-thick gas, as any seed photon from the accretion disk scatters within this medium, ultimately generating black body photons and resulting in the loss of its original information.
Therefore, the gas's state and geometric distribution differ significantly from the typical corona or jet.
More interestingly, the appearance of QPOs corresponds to the formation of a magnetically driven failed wind that might be the source of this Compton-thick gas (\citealp{Miller2020ApJ...904...30M}).
Considering that this gas originates much farther from the black hole, explaining the generation of LFQPO through a geometric progression caused by the Lense-Thirring effect might not be true. Given the previously mentioned direct correlation between the QPO and the Compton-thick magnetically driven failed wind, it is plausible to speculate that the mechanism for generation might arise from perturbations in the magnetic field being transmitted into the failed disk wind.
This process bears a resemblance to the one outlined in the Accretion Ejection Instability (AEI) model (\citealp{Tagger1999A&A...349.1003T}), which elucidates the instability of spiral waves in the density and scale height of a thin disc interlaced with a robust vertical (poloidal) magnetic field.
Moreover, spiral waves not only introduce additional instability but also shape the Rossby vortex at the corotation radius. Subsequently, this vortex induces a twisting of the field lines that traverse this region, leading to the initiation of a vertical Alfven wave. This wave has the ability to transfer both energy and angular momentum from the disk to the corona, thereby playing a crucial role in initiating the formation of a wind and/or jet (\citealp{Ingram2019NewAR..8501524I}).
However, once the mechanism shifts towards being thermally driven, the decoupling of the wind and magnetic fields formed in the outer regions leads to the dissipation of perturbations. Consequently, the QPO also ceases to exist.

\section{Conclusions}

We have observed magnetically driven and thermal-driven transitions during a notable soft X-ray re-brightening in 2021. 
Remarkably, these transitions gave rise to a low-speed 'failed wind' that formed a dense Compton-thick structure. 
During the magnetically driven period, it exhibits elevated ionization and column density levels.
We propose that perturbation within the magnetic field can transmit into the Compton-thick gas and trigger the emergence of LFQPO signals. This shares many similarities with the AEI model.
This unveils a magnetic-related perturbation mechanism hitherto unexplored in the generation of LFQPOs. We eagerly anticipate additional theoretical investigations and observational evidence to substantiate this conjecture.

\begin{acknowledgements}

This research has made use of data obtained from the High Energy Astrophysics Science Archive Research Center (HEASARC), provided by NASA’s Goddard Space Flight Center. This work is grateful for the financial support provided by the Sino-German (CSC-DAAD) Postdoc Scholarship Program, 2022 (57607866). This work is supported by the National Natural Science Foundation of China under grant No. 12173103. This work is also supported by the National Key R\&D Program of China (2021YFA0718500).

\end{acknowledgements}

\bibliographystyle{aa}
\bibliography{ref.bib}

\end{document}